\documentclass[journal]{IEEEtran}
\usepackage{tmi}
\usepackage{cite}
\usepackage{amsmath,amssymb,amsfonts}
\usepackage{algorithmic}
\usepackage{graphicx}
\usepackage{textcomp}

\usepackage{siunitx}
\usepackage{threeparttable}
\usepackage{multirow}
\usepackage{makecell}
\usepackage{booktabs}
\usepackage{float}
\usepackage{gensymb}
\usepackage{hyperref} 
\hypersetup{ hidelinks,
    colorlinks=false,
    allcolors=black,
    pdfstartview=Fit,
    breaklinks=true
}

\def\BibTeX{{\rm B\kern-.05em{\sc i\kern-.025em b}\kern-.08em
    T\kern-.1667em\lower.7ex\hbox{E}\kern-.125emX}}
\begin{document}
% \linenumbers

\title{VibNet: Vibration-Boosted \\Needle Detection in Ultrasound Images
% Invisible Needle Detection in Ultrasound: Leveraging Mechanism Induced Vibration
}
\author{Dianye Huang,  Chenyang Li, Angelos Karlas, Xiangyu Chu\\
K. W. Samuel Au, Nassir Navab, \IEEEmembership{Fellow, IEEE}, Zhongliang Jiang
\thanks{Dianye Huang and Chenyang Li contributed equally to this work. Corresponding authors (Zhongliang Jiang and Xiangyu Chu).}
% \thanks{This paragraph of the first footnote will contain the date on which
% you submitted your paper for review. It will also contain support information,
% including sponsor and financial support acknowledgment. For example, 
% ``This work was supported in part by the U.S. Department of Commerce under Grant BS123456.'' }
\thanks{Dianye Huang, Nassir Navab, and Zhongliang Jiang are with the Chair for Computer Aided Medical Procedures and Augmented Reality (CAMP), Technical University of Munich, Germany, and also with the Munich Center for Machine Learning (MCML), Arcisstraße 21, 80333 Munich, Germany. }
\thanks{Chenyang Li is with the CAMP, Technical University of Munich, Germany.}
\thanks{Xiangyu Chu and K. W. Samuel Au are with the Department of Mechanical and Automation Engineering, The Chinese University of Hong Kong, China, and the Multi-scale Medical Robotics Centre, Hong Kong SAR, China.}
\thanks{Angelos Karlas is with the School of Medicine, Central Institute for Translational Cancer Research (TranslaTUM), Technical University of Munich, Germany.}
}

\maketitle

\begin{abstract}
Precise percutaneous needle detection is crucial for ultrasound (US)-guided interventions. However, inherent limitations such as speckles, needle-like artifacts, and low resolution make it challenging to robustly detect needles, especially when their visibility is reduced or imperceptible. To address this challenge, we propose VibNet, a learning-based framework designed to enhance the robustness and accuracy of needle detection in US images by leveraging periodic vibration applied externally to the needle shafts. VibNet integrates neural Short-Time Fourier Transform and Hough Transform modules to achieve successive sub-goals, including motion feature extraction in the spatiotemporal space, frequency feature aggregation, and needle detection in the Hough space. Due to the periodic subtle vibration, the features are more robust in the frequency domain than in the image intensity domain, making VibNet more effective than traditional intensity-based methods. To demonstrate the effectiveness of VibNet, we conducted experiments on distinct \textit{ex vivo} porcine and bovine tissue samples. The results obtained on porcine samples demonstrate that VibNet effectively detects needles even when their visibility is severely reduced, with a tip error of $1.61\pm1.56~mm$ compared to $8.15\pm9.98~mm$ for UNet and $6.63\pm7.58~mm$ for WNet, and a needle direction error of $1.64\pm1.86^{\circ}$ compared to $9.29\pm15.30^{\circ}$ for UNet and $8.54\pm17.92^{\circ}$ for WNet. Code: https://github.com/marslicy/VibNet.
\end{abstract}


%%%%%%%%%%%%%%%%%% main content
\section{Introduction}
\label{sec:introduction}
\IEEEPARstart{U}{ltrasound}-guided percutaneous needle insertion is widely used in clinical practices~\cite{kuang2016modelling}, such as drug delivery, regional anesthesia, and tissue biopsy. Compared with CT and MRI, ultrasound (US) is real-time and radiation-free. However, inherent limitations such as speckles and artifacts make it challenging to robustly detect and track the percutaneous needle. Furthermore, the bright tissue interfaces in US images can generate needle-like appearing artifacts, complicating needle detection and tracking, especially in deeper regions~\cite{reusz2014needle}. Even when the needle is well-aligned in the 2D US view, poor visibility is still frequently observed (over $50\%$) in practice~\cite{guo2012echogenic}. Thereby, to discern the needle in 2D US images, years of training and experience are required during the insertion process~\cite{jiang2023robotic, bi2024machine}.   
Although the 3D probe has gained increased attention in recent years, its application is still limited due to constraints such as restricted imaging volume, reduced image quality, and higher computational complexity~\cite{jiang2024needle}.

\begin{figure}[t]
    \centering
    \includegraphics[width=0.46\textwidth]{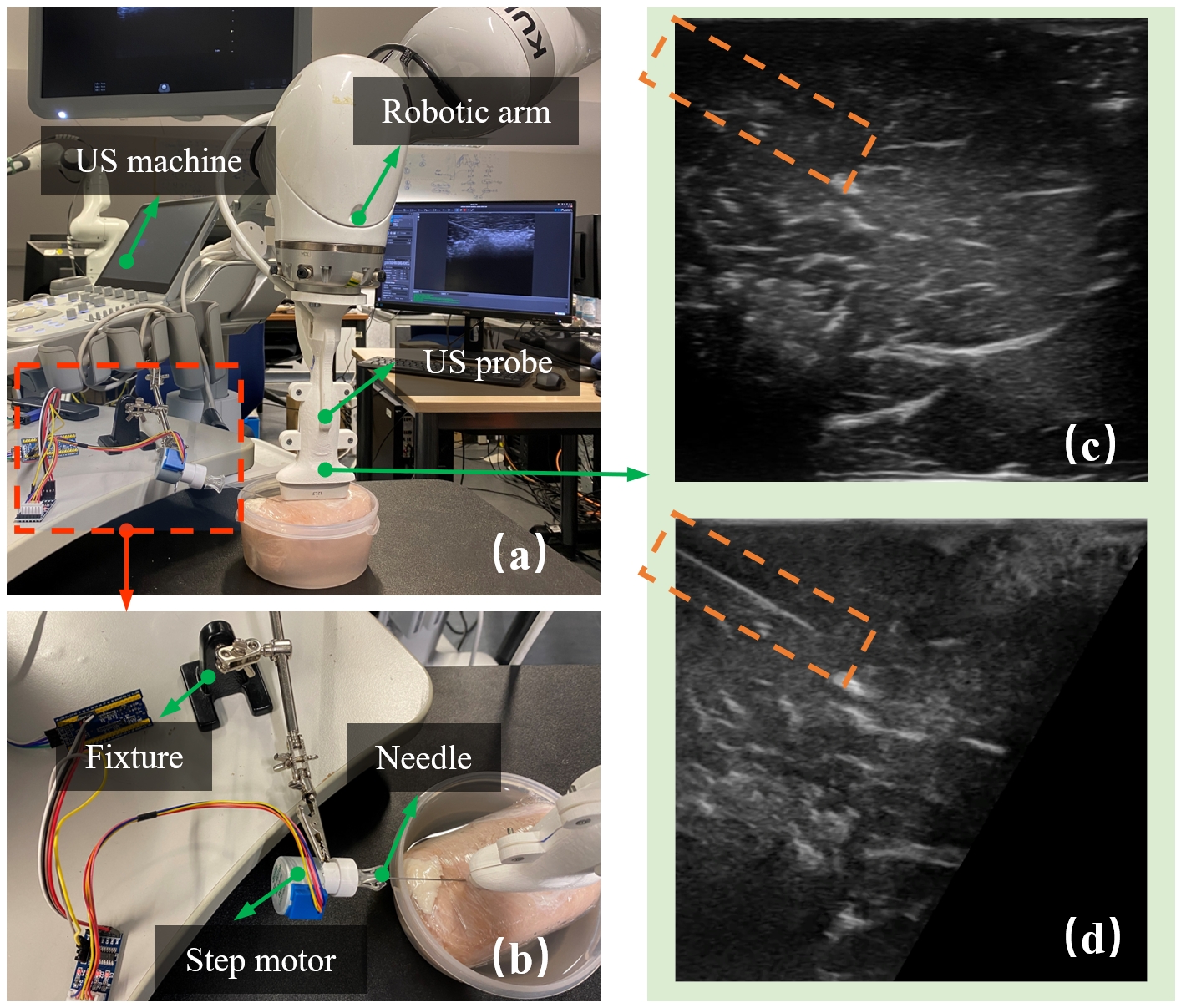}
    \caption{A representative of the \textbf{nearly invisible needle} in US images. (a) and (b) demonstrate data acquisition setup on an \textit{ex-vivo} animal tissue sample. (c) is a typical B-mode image with a nearly invisible needle, and (d) is the B-mode image acquired at the same location with an optimized beam angle, adjusting manually. The dashed orange rectangles indicate the needle's location.}
    \label{fig:introduction}
\end{figure}

Echogenic needles~\cite{hovgesen2022echogenic} have been developed with features like grooves, dimples, or coatings to enhance visibility in US by increasing sound wave reflection or scattering. However, their adoption has been limited due to higher costs, the potential to obscure adjacent anatomical structures with artifacts or bright spots, and variability in effectiveness according to factors like tissue properties, US frequency, etc. Besides, needles can also be modified to produce a color Doppler twinkling signature to help identify it during insertion~\cite{dupere2023new}. However, their bright echogenic surface can induce shadowing artifacts and other distortions on the image, potentially complicating the procedure~\cite{chin2008needle}. To enhance standard needles’ visibility, spatial compounding imaging combines multiple beam-steered US frames~\cite{wiesmann2013compound}. However, this technology can reduce frame rates and needs open access to low-level control to change the beam steering angles, which is not common for most commercial devices used in clinics. Besides, the smoothing effect inherent to the technique may obscure small or subtle features. Active needle-tracking technologies, such as electromagnetic tracking~\cite{schicho2005stability}, provide accurate detection performance. However, they require costly tracking devices, and the integration of the tracking marker at the tip of the standard is also not easy. As a result, there is still a high demand for needle detection based on the standard needle and using B-mode images.

\par
To extract instruments from US images, traditional computer vision and machine learning methods such as the Hough transform~\cite{beigi2021enhancement} and Principal Component Analysis (PCA)~\cite{novotny2003tool} have been investigated. To detect an inserted needle, Kaya~\emph{et al.} applied the Gabor filter on US images, followed by the Random Sample Consensus (RANSAC) line estimator to identify the best-fitted line representing the needle~\cite{kaya2014needle}. However, the performance of such feature-based methods heavily relies on tuning parameters like orientation angles in the Gabor filter.

\par
% CNN based
Recently, benefiting from the remarkable advancements in deep learning, convolutional neural networks (CNNs) have emerged as a pivotal tool in medical image processing~\cite{bi2023mi, huang2023motion, jiang2024intelligent, bi2024synomaly}. A prominent example of this is the U-Net architecture~\cite{ronneberger2015u} and its variants~\cite{jiang2021autonomous, bi2025gaze, jiang2023defcor}, which has been extensively utilized due to its superior performance across a range of medical imaging applications. In regard to instrument segmentation from US images, Gillies~\emph{et al.} trained a U-Net using Dice loss to segment needle-like instruments from various anatomical backgrounds,  achieving overall median errors of $3.5~mm$ for tip localization and $0.8^\circ$ for angle estimation~\cite{gillies2020deep}. To further use the temporal information, Chen~\emph{et al.} proposed a W-Net architecture using two consecutive frames as input~\cite{chen2022automatic}. Alternatively, Yan~\emph{et al.} adopted a transformer module to leverage the temporal information from sequential data~\cite{yan2023learning, yan2024task}. Considering the inserted needle is much smaller than the background, Yang~\emph{et al.} introduce a hybrid loss consisting of contextual loss and focal loss to tackle the problem of imbalanced class distributions~\cite{yang2021efficient}. Tackling the same problem, Mwikirize~\emph{et al.} explicitly used a regional proposal module to encourage the network to effectively detect the region of interest (ROI) and do fine segmentation in the identified ROI, achieving tip and angle errors of $0.23\pm0.05~mm$ and $0.82\pm0.4^\circ$, respectively, over a lumbosacral spine phantom~\cite{mwikirize2018convolution}. To precisely detect the needle tip, Hacihaliloglu~\emph{et al.} proposed an additional search algorithm~\cite{hacihaliloglu2015projection} along the extracted needle shaft. However, the above methods detect the needle by analyzing its appearance, which limits their reliability in scenarios where needle visibility is reduced and the needle-like tissue structures become prominent. To reduce dependence on the spatial texture of the input, one solution is to analyze the temporal variations in pixel intensities across consecutive US images.

Considering the practical problem that the needle shaft may only be partially or even not visible, Mwikirize~\emph{et al.} enhanced the needle tip through digital subtraction of adjacent frames, subsequently using these enhanced images as input for a CNN~\cite{mwikirize2019single}. In their extended work, they introduced a long short-term memory (LSTM)-based fusion module to leverage spatiotemporal information, reducing the tip detection error from $0.74~mm$ to $0.52~mm$~\cite{mwikirize2021time}. Alternatively, to enhance the needle visibility, Cheung~\emph{et al.} steered the US beam to be perpendicular to the inserted needle~\cite{cheung2004enhancement}. The results show the method can significantly improve the needle visibility by maximizing the received reflections. However, the applicability of this method is limited, as most commercial US systems do not provide access to adjust the steering angle during acquisition. In addition, needle vibration has been seen as a promising alternative to enhance needle detection robustness. 
Leveraging the fact that vibration can lead to distinctive color Doppler images, studies have shown that both needle tip and shaft detection accuracy can be improved~\cite{harmat2006needle, jiang2020localization, daoud2022needle, orlando2023power}. Orlando~\emph{et al.}~\cite{orlando2023power} reported achieving an accuracy of $0.4\pm0.2~mm$ on a tissue-equivalent phantom and $0.8\pm0.5~mm$ on five prostate cancer patients by combining Doppler and B-mode images. While these results are promising for prostate applications, performance may decline in procedures involving more complex tissue layers because of the noise in Doppler images. Alternatively, Beigi~\emph{et al.} investigated the potential of applying manual tremor motion assisting the needle detection task under the B-mode images and reported that it could work better for convex probes because the initial portion of the shaft is usually detectable near the puncture site~\cite{beigi2016spectral}. However, the performance was witnessed to be decreased when testing on in-vivo tissue due to the noisy motion of surrounding tissues. To mitigate the negative impact caused by the subtle tremor motion of a hand-held US probe, Beigi~\emph{et al.} proposed a tracking framework using spatiotemporal features~\cite{beigi2017detection}. However, the final performance heavily relies on customized features, and the computation cost is high (only one frame per second) because of the conventional multi-scale spatial decomposition. 

To address these challenges, this study proposed a deep-learning framework, VibNet, aiming to enhance both the efficiency and accuracy of the needle segmentation from US images by leveraging vibration applied externally on the needle shaft. By detecting the periodic vibration pattern applied by an external motorized mechanism in the frequency domain, the proposed method mitigates the dependence on image intensity. Therefore, it can maintain its effectiveness even when the inserted needle becomes imperceptible in US images. Since the needle is firmly covered by tissue, the subtle eccentric  motion causes no additional damage compared to the manual pull-and-push motion commonly used in current practice. To the best of our knowledge, this is the first end-to-end deep learning framework that leverages vibration information for needle detection. The proposed VibNet consists of three modules: temporal feature extraction, frequency feature aggregation, and needle detection. The main contributions are summarized as follows:
\begin{enumerate}
  \item The first end-to-end VibNet, featuring temporal and frequency feature extraction along with needle prediction modules, is introduced to leverage periodic needle vibrations for robust detection, especially in cases where needle visibility is reduced or nearly invisible.

  \item The discrete Short-Time Fourier Transform (STFT) is approximated using 1D convolution with initialized kernels to extract the time-frequency spectrogram from the temporal signal generated by image intensity at each pixel across a sequence of consecutive images.
  
  \item The deep Hough transformation is applied to latent features to enhance needle tip detection accuracy by mitigating severe class imbalance in B-mode image space, where a single point corresponds to a sinusoidal-shaped curve in Hough space.
  %---$>$ 1) parameters, class imbalance

  % \item add more things if necessary
\end{enumerate}
To validate the effectiveness of the proposed VibNet, cross-validation experiments were performed on two  \textit{ex vivo} tissue samples. The results demonstrate that both robustness and detection accuracy are enhanced, even when the needle becomes visually invisible in extreme cases. Regarding the use case, vibration can be introduced when needle visibility is reduced in US images, eliminating the need for manual pull-and-push adjustments commonly used in current practice.

\par
The rest of this paper is structured as follows: Section \ref{sec:method} elaborates on the design of the proposed detection framework. Experimental results on two different \textit{ex vivo} animal tissue samples are presented and analyzed in Section \ref{sec:exp}. The discussion and conclusion are in Sections~\ref{sec:discussion} and \ref{sec:conclusion}, respectively. 

\section{Method}
\label{sec:method}
\par
This section will elaborate on each module in the proposed VibNet. The goal of VibNet is to leverage the known periodic vibration pattern to make it possible to detect the needle shaft and tip when the inserted needle becomes less visible or nearly invisible in the US image. This capability is important in clinical practice because it can extend the human ability to identify challenging needles using the naked eye. In contrast to the conventional needle segmentation studies that rely on image intensity, VibNet is developed based on the features captured in the frequency domain. This enables VibNet to detect needles with poor visibility because the motion pattern of the pixels belonging to the vibrating needle is distinctive from that of the background pixels. 

\begin{figure}[!h]
    \centering
    \includegraphics[width=0.485\textwidth]{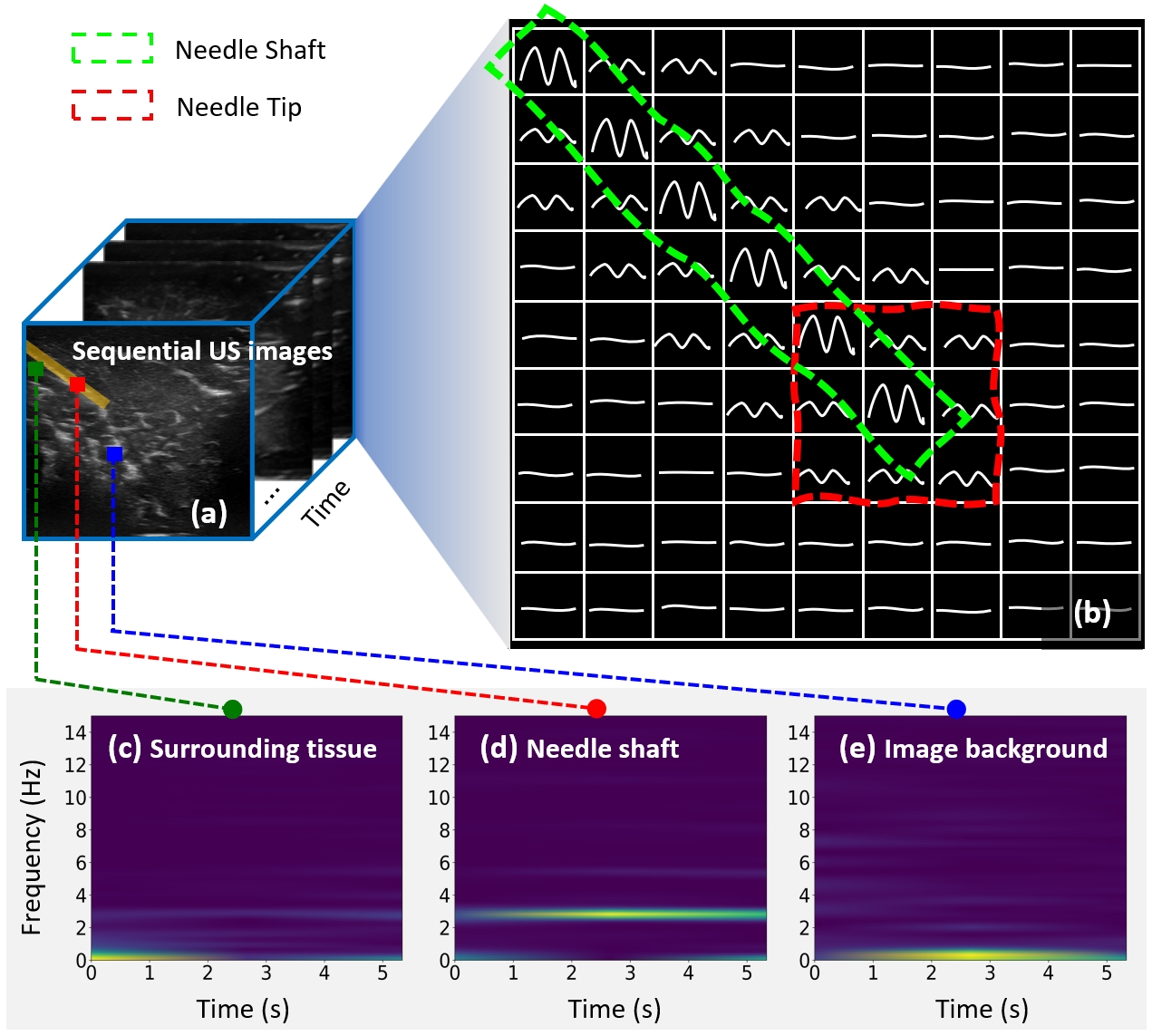}
    \caption{An intuitive explanation of the philosophy of the proposed method. (a). presents three image patches selected for analysis. (b). The vibrated needle induces different motion patterns in the US image sequence, making the needle itself perceptible to the network. It is noteworthy that the sinusoidal patterns were manually created to represent the ideal variations in vibration frequency and amplitude at each pixel. (c)-(e) are computed corresponding spectrograms of the three patches, demonstrating the distinct motion patterns. 
    }
    \label{fig:intuitive}
\end{figure}

\par
Fig. \ref{fig:intuitive} intuitively presents the underlying philosophy of the proposed method. Three sampling patches depicted in red, green, and blue dots are obtained from the vibrating needle shaft, surrounding tissue, and image background, respectively. It can be seen from Figs.~\ref{fig:intuitive}~(c), (d), and (e) that the patch on the shaft has a visible and dominating main frequency (around $2.5~Hz$) during the vibration, while the background does not show any dominating frequency. Since surrounding tissues will be moved together with the needle, a dominating frequency is also witnessed at the same level as the one in Fig.~\ref{fig:intuitive}~(c). However, the magnitude is significantly smaller, as demonstrated by the much lighter intensity of the line representing the vibrating frequency. This observation indicates that the needle shaft and tip can be detected by distinguishing the different feature appearances of individual pixels in the frequency domain. As illustrated in Fig.~\ref{fig:intuitive}~(b) in ideal case, the pixels exhibiting a clear dominant frequency and consistent amplitude will distribute along a line (highlighted by the green rectangle) and can be considered as the inserted needle. 

\par
\begin{figure*}[!t]
    \centering
    \includegraphics[width=0.90\textwidth]{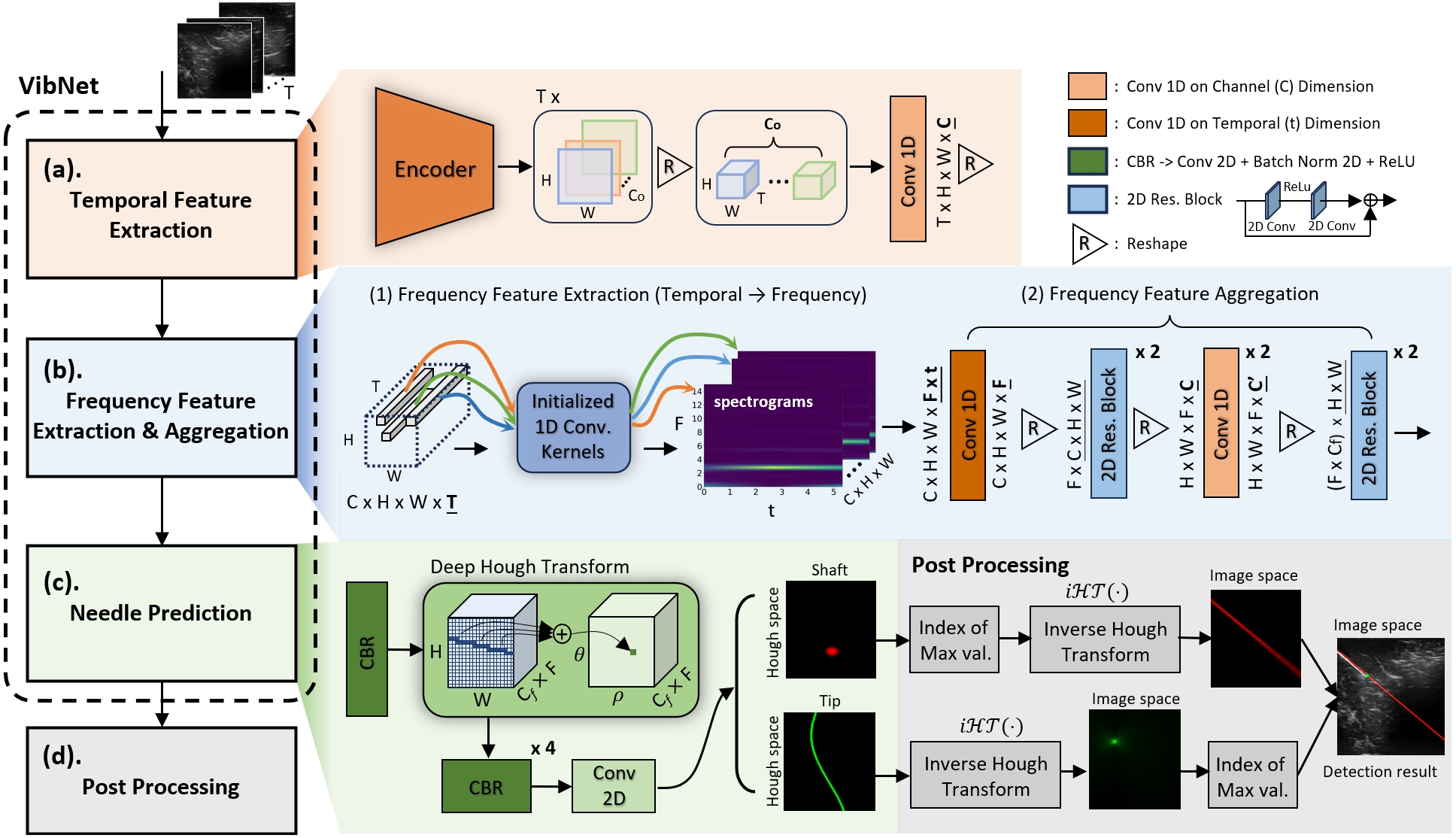}
    \caption{Overview of needle detection pipeline. VibNet comprises modules for (a). Temporal feature extraction in the spatio-temporal space, (b). Frequency feature extraction and aggregation in Frequency space, and (c). Needle shaft and tip prediction in Hough space. The needle's location is finally determined by (d). Post-processing the output images from VibNet. 
    For further details of module (d), one can refer to \textit{Section} \ref{sec:method}\textit{-C 2)}.
    %\todo{showing the whole pipeline.}
    }
    \label{fig:network_overview}
\end{figure*}

To avoid causing additional trauma, a subtle vibration is generated through periodic eccentric motion, with the needle center fixed at a small offset from the motor's rotation center. To extract the frequency features from such subtle vibration in the whole image space, we utilize a pre-trained encoder originally designed to amplify minor video motions from consecutive frames. Then, the temporal data (pixel-wise intensity value in the identical pixel location of a set of US image inputs) is transferred into the frequency domain and aggregated in the intermediate feature space. To tackle the challenging problem of needle tip localization, the frequency features are further transferred into Hough space using deep Hough transform. The overall structure of VibNet is depicted in Fig.~\ref{fig:network_overview}. A series of consecutive US images are used as input, and the output consists of two-channel images, which indicate the needle shaft and tip in Hough space. The details of VibNet's three main modules—(a) temporal feature extraction, (b) frequency feature extraction and aggregation, and (c) needle prediction—are presented in the following subsections. 

\subsection{Vibration Encoding from Sequential US Images}
\par
To characterize the subtle vibration of the needle, sequential US images $\mathbf{I}\in\mathbb{R}^{H\times W}$ are used as inputs. Instead of using customized decomposition filters, such as complex steerable filters~\cite{wadhwa2013phase}, a CNN-based encoder is applied here to extract the representation in order to mitigate the complexity of hand-craft filters and to enhance the time efficiency. It was reported in~\cite{wadhwa2013phase} that processing a $512 \times 512$ video with 300 frames using an octave-bandwidth pyramid takes $56~s$. Compared with customized feature designs, a CNN-based process can enhance both robustness to noise and computational efficiency~\cite{oh2018learning}. To improve the sensitivity for subtle motions in consecutive images, a pre-trained encoder $E_{shape}(\mathbf{I})$ presented by Oh~\emph{et al.}~\cite{oh2018learning} based on extensive synthetic data with moving objects is used. This encoder has been proven to be able to characterize the subtle motion of objects for video magnification. Repeating using this encoder for $T$ images, a feature representation $\mathbf{S}\in\mathbb{R}^{T\times C_o\times H\times W}$ for sequential data can be computed. Then, we apply a 1D convolution, $\textbf{S}_t = Conv_{1d}(\mathbf{S})$, in the dimension of the feature channel to compress the number of features $(C_o\rightarrow C)$ while maintaining spatial-temporal correspondence to original inputs. The output of the vibration encoding module is denoted as $(\textbf{S}_t)$ in the temporal dimension, with dimensions $(C \times H \times W \times \underline{T})$.

\subsection{Feature Aggregation}
\subsubsection{Domain Transformation} Specific to thin objects like needles, their visibility is severely affected by the US speckle noise, tissue artifacts, etc. Unlike the image intensity, the periodic vibration pattern in the frequency domain is more robust and distinctive (see Fig.~\ref{fig:intuitive}). In order to extract the frequency characteristic from temporal data, Discrete Fourier Transform (DFT) is a commonly used method. The domain transfer process is defined as follows:
\begin{equation}
        \hat{\mathbf{x}}_{[k]} = \sum_{n=0}^{N-1}\mathbf{x}_{[n]} e^{-j2\pi k\frac{n}{N}} 
        = \mathbf{x}\cdot\mathbf{b}_{cos}^{k}+ j\mathbf{x}\cdot\mathbf{b}_{sin}^{k}
    \label{eq-dft}
\end{equation}
where $N$ is the length of the input signal $\mathbf{x}\in\mathbb{R}^{N}$ indexed by $n$; $\hat{\mathbf{x}}_{[k]}$ denotes the $k$-th spectrum of the output signal $\hat{\mathbf{x}}$, corresponding to the frequency $f_k=\frac{k}{N}f_s$, where $f_s$ is the sampling frequency. According to the Shannon Sampling Theorem, feasible $f_k$ is halved of $f_s$, namely, $k = 0, 1, \cdots, \lfloor N/2\rfloor$. The symbol $\lfloor \rfloor$ represents the floor function. Therefore, we have complex variable $\hat{\mathbf{x}}\in\mathbb{C}^{\lfloor N/2\rfloor}$. The symbol ``$\cdot$'' stands for the dot-product of two vectors. The vectors $\mathbf{b}_{cos}^{k}$ and $\mathbf{b}_{sin}^{k}$ represent the $k$-th basis functions for DFT, sampled at $\{n/N\}_{n\in\{1,2,\dots,N-1\}}$. 
\begin{equation*}
    \begin{aligned}
        \mathbf{b}_{cos}^{k}=&\left\{cos\Big(2\pi k n/N\Big)\right\},~n\in\{0,1,\dots,N-1\} \\
        \mathbf{b}_{sin}^{k}=&\left\{-sin\Big(2\pi k n/N\Big)\right\},~n\in\{0,1,\dots,N-1\}
    \end{aligned}
    \label{eq:ft_weights}
\end{equation*}

Based on (\ref{eq-dft}), to decompose 1D temporal data $\mathbf{x}$ into different frequency ranges, we can also compute the dot-product between $\mathbf{x}$ and the row vectors of a matrix $\mathbf{W}$. By doing so, $\mathbf{\hat{x}}$ can be represented as $\mathbf{x}_f \in \mathbb{R}^{2\cdot\lfloor N/2\rfloor}$ consisting of real numbers.
\begin{equation}
    \begin{aligned}
        \mathbf{x}_f =& \mathbf{x}\cdot\mathbf{W}
        \Leftrightarrow
        \left[
        \begin{array}{c}
        \vdots \\
        \mathrm{Real}(\mathbf{\hat{x}_{[k]}}) \\ 
        \mathrm{Img}~(\mathbf{\hat{x}_{[k]}}) \\ 
        \vdots \\
        \end{array}
        \right] = 
        \left[
        \begin{array}{c}
        \vdots \\
        \mathbf{x}\cdot\mathbf{b}^k_{cos} \\ 
        \mathbf{x}\cdot\mathbf{b}^k_{sin} \\ 
        \vdots \\
        \end{array}
        \right]
        \\
    \end{aligned}
    \label{eq:conv_ft}
\end{equation}
Since $\mathbf{\hat{x}}$ is a complex vector, similar to the Fourier-initialized convolution (FIC) layer in \cite{li2021units}, each spectral $\mathbf{\hat{x}}_{[k]}$ is computed using two vectors: one for the real part and one for the imaginary part. Thus, $\mathbf{W}\in\mathbb{R}^{2\cdot\lfloor N/2\rfloor\times N}$, comprising $\lfloor N/2\rfloor$ pairs of cosine and sine components $\left\{\mathbf{b}_{cos}^k,~\mathbf{b}_{sin}^k \right\}_{k\in\left\{0,1,\dots,\lfloor N/2\rfloor-1\right\}}$, is defined as:
\begin{equation}
    \begin{aligned}
        \mathbf{W}_{\left[2k, n\right]}&=\mathbf{b}_{cos[n]}^{k}=cos\left(2 \pi nk/N\right)\\
        \mathbf{W}_{\left[2k+1, n\right]}&=\mathbf{b}_{sin[n]}^{k}=-sin\left(2 \pi nk/N\right)\\
    \end{aligned}
    \label{eq:w_init}
\end{equation}

\par 
In order to capture the spectral content of a signal that evolves over time, a discrete Short-Time-Fourier-Transform (STFT) is used in this study. In STFT, a moving window function $w(\cdot)$ of length $N_w$ is applied to the input signal $\mathbf{y}\in\mathbb{R}^{L}$ to maintain temporal information while computing the time-frequency spectrogram $\hat{\mathbf{Y}}$. In this study, $N_w$ is empirically selected to 10. DFT is a special case when $N_w=L$ and $w(\cdot)=1$. In other words, STFT is the result computed by repeating the DFT process for each moving window on the input signal. The detailed definition is displayed as follows:
\begin{equation}
\begin{aligned}
    \mathbf{\hat{Y}}_{\left[k, m\right]} =& \sum_{n=0}^{N_w-1}\mathbf{y}_{[n+mH]}w(n)e^{-j2\pi k\frac{n}{N_w}}\\    
    =&\Big(\mathbf{\Bar{y}}_{m:} \odot \mathbf{w} \Big)\cdot\mathbf{b}_{cos}^{k}+j\Big(\mathbf{\bar{y}}_{m:} \odot \mathbf{w} \Big) 
 \cdot\mathbf{b}_{sin}^{k}\\
\end{aligned}
\label{eq:stft}
\end{equation}
where $\odot$ denotes element-wise product of two vectors; $\mathbf{w}=\{w(n)\}_{n\in\{0,\cdots,N_w-1\}}$; $k$ and $m$ are the frequency and temporal indices of the spectrogram $\mathbf{\hat{Y}}\in\mathbb{R}^{\lfloor N/2 \rfloor\times M}$, where $M=\lfloor(L-N)/H\rfloor$; $H$ is the stride length; $\mathbf{\bar{y}}_{m:}\in\mathbb{R}^{N_w}$ denotes a segment of $\mathbf{y}$ starting from index $mH$ up to $mH+N_w-1$. In this study, the rectangle window function is used. To have the best temporal resolution, $H$ is set to one in this study. The STFT given in (\ref{eq:stft}) is equivalent to applying a 1D convolution with $k$ kernels to a 1D signal, where the stride size is $H$. When combined with the computation of DFT in (\ref{eq:conv_ft}), equation (\ref{eq:stft}) can be rewritten as follows:
\begin{equation}
    \begin{aligned}
    % \mathbf{Y}_f&=\mathbf{y}\ast{\scriptsize{\mathrm{Conv}}_{1d}}(\mathbf{W_{stft}})\\
    \mathbf{Y}_f&=\mathbf{y}\ast{\mathrm{Conv}}_{1d}(\mathbf{W_{stft}})\\
    \end{aligned}
    \label{eq:conv_stft}
\end{equation}
where ``$\ast$" denotes the convolution operator; the 1D convolution kernels are initialized as $\mathbf{W_{stft}}=\mathbf{W}$ defined in (\ref{eq:w_init}), because $w(\cdot)$ here is a rectangular function.

\par
Let $\mathbf{z}\in\mathbb{R}^{T}$ ($T>N_w$) be a 1D signal along the temporal dimension of $\mathbf{S}_t\in\mathbb{R}^{C \times H \times W \times T}$. After applying Conv$_{1d}(\mathbf{W})$ to each individual temporal series input, a set of 2D spectrograms $\mathbf{Z}_{f}\in\mathbb{R}^{F\times t}$ described in both frequency $F$ and  temporal $t$ dimensions are obtained. To reduce the dimension of the STFT module, another 1D convolution is applied to reduce the feature size from $(C\times H \times W \times \underline{F \times t})$ to $(C\times H \times W \times \underline{F})$. 

\par
\subsubsection{Frequency Feature Aggregation} To further extract features representing different motion patterns, we aggregate the multi-channel feature map by successively applying convolutions on the spatial dimension using ResBlock~\cite{he2016deep} and on the channel dimension using 1D convolution. The identical ResBlock layer is repeated twice to enhance feature extraction. Subsequently, the latent features are fused by successively reducing the number of channels using two 1D convolution layers. This process ensures that the latent features maintain the same resolution as the input images, enabling us to trace back the spatial information to the original inputs. The frequency feature aggregation module concludes with a 2D ResBlock, yielding an aggregated feature $\mathbf{F}_f\in\mathbb{R}^{(C_f\cdot F)\times H\times W}$.

\subsection{Needle Shaft and Tip Localization}
\subsubsection{Needle Prediction}
\par
For US-guided interventions, close monitoring of the inserted needle tip is crucial to ensure the safety and precise targeting of the region of interest. To this end, we transfer the intermediate feature into Hough space. In this way, the detection of a single point in image space has been cast into the problem of detecting a curve in Hough space (see Fig.~\ref{fig:network_overview}). By increasing the number of pixels in the ground truth annotation of the needle tip, the trained model can be more robust for tip detection and mitigate the negative effect caused by the non-perfect annotation of the needle tip. Additionally, by supervising both the needle shaft and tip features in Hough space, VibNet is constrained to detect the line-like needle geometry rather than a flashing pixel with a frequency similar to the needle vibration.

\par
To apply the Hough Transform in latent space, the Deep Hough transform (DHT) module proposed in~\cite{zhao2021deep} is applied on latent feature $\mathbf{F}_f\in\mathbb{R}^{(C_f\cdot F)\times H\times W}$. To reduce the processing time of DHT, the latent features pass through a 2D convolutional block (Conv+BN+ReLU) that compresses the number of channels before they are processed by a DHT module. To further aggregate the features spatially, we employed four 2D convolutional blocks after applying DHT. This is followed by another 2D convolution with $1\times 1$ kernel to generate a two-channel prediction image in Hough space. The shaft is presented as a single point in the first channel $\mathbf{I}_p^{1}$, while the tip is presented as a sinusoidal-shaped wave in the second channel $\mathbf{I}_p^{2}$. The entire network was trained in a supervised manner, using the ground truth of both shaft and tip in Hough space. Details regarding the loss design and training of the VibNet can be found in Sec.~\ref{sec:II_loss}. 

\subsubsection{Post Processing}
\par
In order to extract the needle shaft and tip in image space, the 2-channel prediction $\mathbf{I}_p$ in Hough space is further processed as follows. 
\begin{equation}
\begin{aligned}
    &(\theta_s,\rho_s)=\mathop{\mathrm{arg\,max}_{\theta,\rho}}\mathbf{I}_{p}^{1} \\
    &(x_t,y_t)=\mathop{\mathrm{arg\,max}_{x,y}}\sum_{\theta_i, \rho_i\in\Theta_{\mathbf{I}_p^2}}i\mathcal{HT}\Big({I}_p^2({\theta_i, \rho_i}),\theta_i, \rho_i\Big)
\end{aligned}
\label{eq:post_process}
\end{equation}
where each pixel intensity in $\mathbf{I}_p^{1}$ represents the probability of the current pixel being the target shaft location; while in $\mathbf{I}_p^{2}$, it represents the probability of its corresponding line in the image space intersecting with the target tip location; $i\mathcal{HT}(\cdot)$ denotes the probability-weighted inverse Hough transform operation, ${I}_{p}^{2}({\theta_i, \rho_i})$ is the pixel intensity of $\mathbf{I}_{p}^{2}$ indexed by $\theta_i, \rho_i$, and $\Theta_{\mathbf{I}_p^{2}}$ is a collection of the pixel positions corresponding to top $p\%$ pixel intensities in $\mathbf{I}_p^{2}$. %\todo{indicate the value of p\%}

The pixel with the highest intensity in $\mathbf{I}_p^{1}$ determines the shaft location $(\theta_s,~\rho_s)$. For the tip detection, all elements in $\Theta_{\mathbf{I}_p^2}$ are selected and processed using $i\mathcal{H}\mathcal{T}(\cdot)$. This operation projects the pixel coordinates back to the image space, and the resulting images are then weighted by the corresponding pixel intensities in the Hough space and summed up to generate a new image. Finally, the pixel with the highest intensity in this image determines the tip location $(x_t,~ y_t)$. An intuitive illustration of the post-processing procedures is depicted in the last module of Fig. \ref{fig:network_overview}. 

\subsection{Training Loss} ~\label{sec:II_loss}
\par
Since the needle is a thin object, the number of the corresponding pixels is much lower than the pixels of the image background. This will cast the issue of class imbalance. Therefore, the focal loss is used for needle segmentation in this study. In addition, considering the needle shaft will turn into a signal pixel in Hough space, a probabilistic bluer process is carried out to mitigate the negative effect caused by non-perfect annotation.

\subsubsection{Gaussian-Blurred Ground Truth}
We render a Gaussian-shaped peak into the shaft ground truth images with a 2D Gaussian kernel and utilize a 1D Gaussian kernel to blur the tip ground truth image row-by-row. 
\begin{equation}
    \mathcal{G}\Big((\theta_t,~\rho_t), \sigma\Big)=\exp \left(-\frac{\left(\theta-\theta_t\right)^2+\left(\rho-\rho_t\right)^2}{2 \sigma^2}\right)
\end{equation}
where $\sigma$, $(\theta_t,~\rho_t)$ denote the standard deviation and the ground truth location of the needle shaft in the Hough space. These Gaussian kernels assign Gaussian distributed probabilities to pixels adjacent to the ground truth, thereby taking inevitable labeling errors into account. Moreover, the learning of the model can be accelerated, as the neighborhood of the ground truth may have similar motion patterns with less magnitude. 

\subsubsection{Focal Loss Function}
Instead of using the standard Binary Cross-Entropy (BCE) loss, we employ Focal Loss~\cite{lin2017focal, zhou2020tracking} in this study. It can increase the relative weight for challenging samples whose predictions significantly differ from the ground truth. 
\begin{equation}
L=-\frac{1}{N_p} \sum_{\theta,\rho}\left\{
\begin{array}{l}
\left(1-\hat{Y}_{\theta,\rho}\right)^\alpha \log \left(\hat{Y}_{\theta,\rho}\right)~~~~~~~~~\text{if }Y_{\theta,\rho}=1 \\
\left(1-Y_{\theta,\rho}\right)^\beta\left(\hat{Y}_{\theta,\rho}\right)^\alpha\log \left(1-\hat{Y}_{\theta,\rho}\right)~\text{others} 
\end{array}\right.
\label{eq:focal_loss}
\end{equation}
where $\hat{Y}_{\theta,\rho}$ is the prediction result, $Y_{\theta,\rho}$ is the Guassian-blurred ground truth, $\alpha$ and $\beta$ are hyper-parameters of the Focal Loss, and $N_p$ is the number of pixels in a Hough space image. The final hybrid loss is structured as follows: 
\begin{equation}
    L_{total} = \gamma L_{shaft} + (1-\gamma)L_{tip}
\end{equation}
where $L_{shaft}$ and $L_{tip}$ are the Focal Loss presented in (\ref{eq:focal_loss}), and $\gamma$ is a hyper-parameter that balances the shaft and tip prediction losses. Based on the experiments, $\gamma$ is empirically determined as $0.95$ here. 

\section{Experiments and results}
\label{sec:exp}
\subsection{Experimental Setup}
\begin{figure}[!h]
    \centering
    \includegraphics[width=0.48\textwidth]{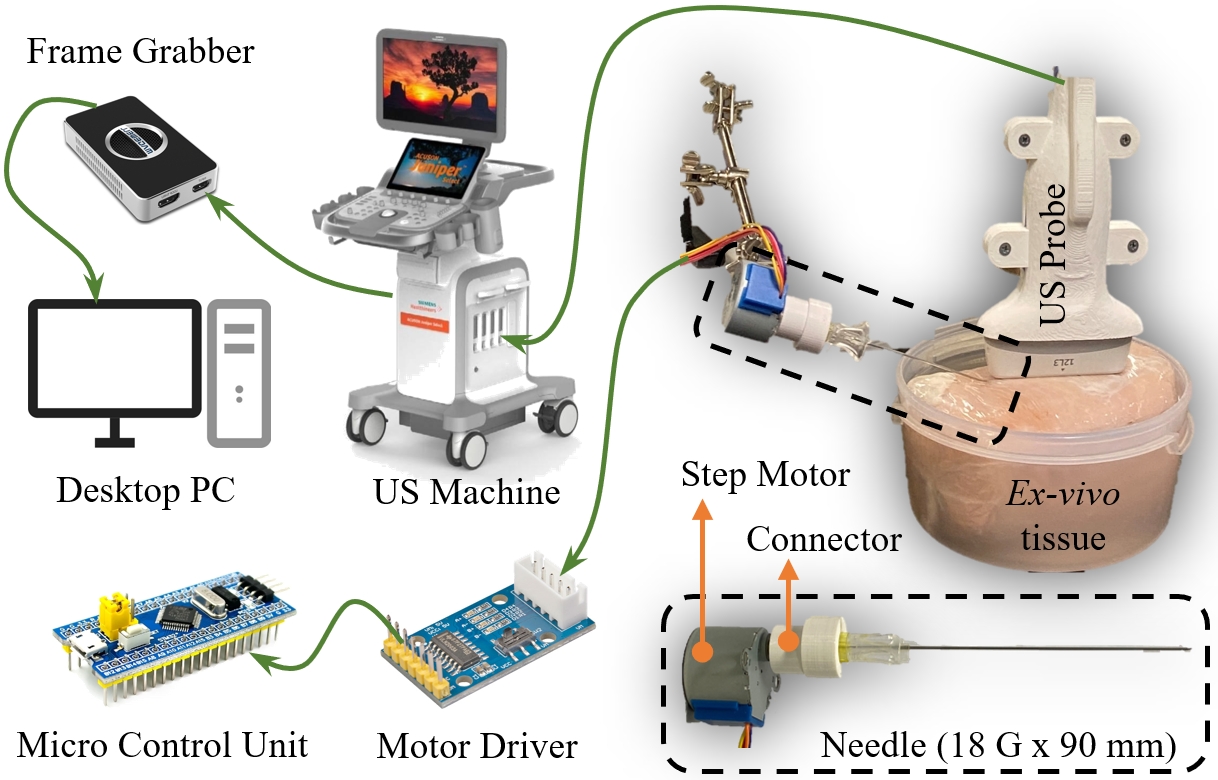}
    \caption{Hardware setup and connections for data acquisition.}
    \label{fig:experiment_settings}
\end{figure}

The experimental setup is shown in Fig.~\ref{fig:experiment_settings}. The US images were captured at $30~fps$ using a linear US probe (12L3, footprint: 51.3 mm) connected to the Siemens Juniper US Machine (ACUSON Juniper, SIEMENS AG, Germany). Then, the images were streamed to a desktop PC via a frame grabber (MAGEWELL, China). To induce needle vibration, a 3D-printed eccentric connector is utilized to link the motor shaft to the needle hub, where the offset is only 1.0 mm between the motor center and needle center. The inserted needle ($18~G$, length: $90~mm$) was slightly vibrated at approximately $2.5~Hz$ by a step motor (28BYJ-48-5V, Pollin Electronic GmbH, Germany). This rigid needle is designed for facet joint injections to diagnose and manage pain with less bending. The step motor is controlled by a Microcontroller Unit (STM32F103C8T6, STMicroelectronics NV, Switzerland) to periodically rotate $15.8~deg$ clockwise and then $15.8~deg$ counterclockwise. The resulting needle vibration amplitude at the motor assembly is $0.28$ mm. The training was performed on a workstation with one GPU (Nvidia GeForce RTX 4070) and one CPU (Intel$^\circledR$ Core\texttrademark ~i7-13700KF Processor). To mitigate model overfitting, we used early stopping during training, and augmented the input image sequence with horizontal flipping, Gaussian blurring, and adjustments of image contrast and brightness.  The model was trained using an Adam optimizer with learning rate 1e-4, betas (0.9, 0.999). 

\subsection{Data Preparation} 
\begin{figure}[!h]
    \centering
    \includegraphics[width=0.485\textwidth]{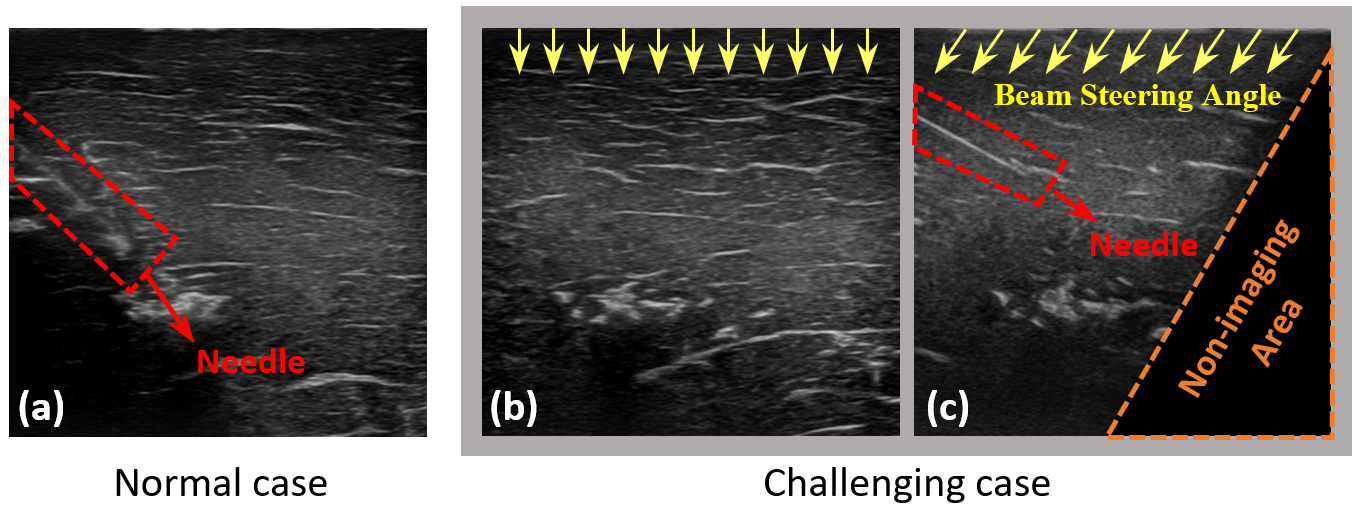}
    \caption{An illustration of (a) a normal image with good needle visibility, (b) and (c) show a challenging case where US images are obtained at the same imaging plane with and without an optimal beam steering angle adjustment, respectively. The insertion angle for the normal case and challenging cases are $45^{\circ}$ and $30^{\circ}$, respectively.}
    \label{fig:usimage_visibility}
\end{figure}

A total of 417 ten-second US videos (roughly 125K frames) were recorded, comprising 206 and 211 videos from \emph{ex vivo} porcine (4 pieces) and bovine (3 pieces) tissue samples, respectively. Before data acquisition, the needle was manually inserted into the ex-vivo tissue, and the needle holder was fixed. The US beam angle was then adjusted to obtain optimal needle visualization [see Fig. \ref{fig:usimage_visibility} (c)]. The video was recorded, consisting of a 3-second Clip A with optimally adjusted beam angle for labeling and a 7-second Clip B that records the vibration of the inserted needle without beam angle adjustment (could be visually invisible). To increase the diversity of the dataset, the imaging location was randomly selected, and the videos were captured at varying angles (ranging from 20° to 50°) and depths (ranging from 10 mm to 40 mm). The annotation was done by an experienced US user under the guidance of a doctor on the images with an optimized steering angle. The training and test sets are initially divided by videos, and then multiple sequences are extracted from Clip B accordingly. Each sequence has $T=30$ frames with a resolution of $328 \times 335\ (H \times W)$.

Table \ref{tab:dataset} provides an overview of the dataset in this study. For each type of \textit{ex-vivo} tissue sample (Bovine and Porcine), we divided the dataset into training (64\%), validation(16\%), and test (20\%) sets. To systematically analyze the impact caused by varying needle visibility, the test set consisting of 40 videos was further divided into normal cases with good visibility (20 videos) and challenging cases with less visibility, respectively. An intuitive visualization of the challenge image and normal image obtained on porcine data has been shown in Fig.~\ref{fig:usimage_visibility}. In the normal case, the needle can be recognized, while in the challenging case [see Fig.~\ref{fig:usimage_visibility}~(b)], it is nearly invisible. To annotate the challenging image set, we first manually adjust the US beam steering angle to enhance their visibility [see Fig.~\ref{fig:usimage_visibility}~(c)]. It is important to note that in most commercial US devices, the steering angle can only be adjusted offline. The enhanced visibility images are utilized solely to obtain ground truth in challenging scenarios. Additionally, to analyze segmentation performance based on the insertion angle, the test set is subdivided into different categories according to the insertion angle. Insertions with angles greater than $30^{\circ}$ are categorized as steep insertions, while those with angles equal to or less than $30^{\circ}$ are considered low-angle insertions. Low-angle insertion is commonly used for superficial structures, such as breast or skin biopsies, while steep-angle insertion is often required for regional anesthesia and deep-structure biopsies, such as the liver and kidney. Steep insertion usually can result in reduced visibility due to anisotropic reflection compared to low-angle insertion. However, the final appearance is highly dependent on coupled factors, including depth, tissue structure, and US noise. Representative examples are shown in Fig.~\ref{fig:usimage_visibility}~(a) and (b), where the normal case involves a larger insertion angle than the challenging example. Similar observations were reported in~\cite{souza2018ultrasound}. In addition, only porcine data is used to validate the impact of insertion angle due to its better contrast and brightness from rich fat layers, helping to eliminate the coupling effects of low brightness in bovine data. To further rule out the effect of visibility coupling, the steep and low-angle insertion sets are divided based on the normal set. It is noteworthy that one data point in Pork$_{norm}$ was obtained at an insertion angle of $29.39^{\circ}$. Since this value is too close to the threshold and could blur the distinction between low-angle and steep insertion, it was excluded from Pork$_{shw}$. 

\begin{table}[!h]
\caption{Dataset division for experiments.}
\label{tab:dataset}
\centering
\tabcolsep=0.20cm
\resizebox{0.485\textwidth}{!}{%
\begin{tabular}{llll}
\toprule
\multicolumn{2}{c}{\textbf{Data}} & \textbf{Porcine Tissue Sample} & \textbf{Bovine Tissue Sample} \\
\midrule
\multicolumn{2}{c}{Training Set} & \scriptsize{$\mathrm{Pork_{train}}$}, 166 (33200) & \scriptsize{$\mathrm{Beef_{train}}$}, 171 (35940) \\
\midrule
\multirow{6}{*}{Test Set}  & \scriptsize{all}   & \scriptsize{$\mathrm{Pork_{test}}$}, 40 (7878) & \scriptsize{$\mathrm{Beef_{test}}$}, 40 (8327) \\ 
% \midrule
\cmidrule{2-4}
& \scriptsize{Normal} & \scriptsize{$\mathrm{Pork_{norm}}$}, 20 (3855) & \scriptsize{$\mathrm{Beef_{norm}}$}, 20 (4191) \\
& \scriptsize{Challenging} & \scriptsize{$\mathrm{Pork_{chal}}$}, 20 (4023) & \scriptsize{$\mathrm{Beef_{chal}}$}, 20 (4136) \\ 
% \midrule
\cmidrule{2-4}
& \scriptsize{Angle$\leq$30°} & \scriptsize{$\mathrm{Pork_{shw}}$}, 9 (1760) & 
\scriptsize{$\mathrm{Beef_{shw}}$}, 10 (2101) \\
% ~~~~~~~~~~~$\backslash$ \\
& \scriptsize{Angle$>$30°} & \scriptsize{$\mathrm{Pork_{stp}}$}, 10 (1888) & 
\scriptsize{$\mathrm{Beef_{stp}}$}, 10 (2090) \\
% ~~~~~~~~~~~$\backslash$ \\ 
% \midrule
\bottomrule
\multicolumn{4}{l}{\scriptsize The numbers are described in the following format: \#num of videos (\#num of sequences)}
\end{tabular}
}
\end{table}

\begin{table*}
\caption{Needle Detection Performance Comparison Based on Needle Visibility}
\label{tab:comparison_seen}
\centering
\tabcolsep=0.20cm
\resizebox{0.90\textwidth}{!}{%
\begin{tabular}{lcccc|cccc}
\toprule
% Method & Test Dataset & Angle Error (°) & Tip Error (mm) & Outlier (\%) \\
\multirow{2}{*}{\textbf{Methods}} & 
% \multirow{2}{*}{\scriptsize{\textbf{Test Set}}{\Large$\backslash$}\scriptsize{\textbf{Training}}} 
\multirow{2}{*}{\textbf{Test Set}}
& \multicolumn{3}{c}{\textbf{Porcine Training Set,} \scriptsize{$\mathrm{Pork_{train}}$}} & 
% \multirow{2}{*}{\scriptsize{\textbf{Test Set}}{\Large$\backslash$}\scriptsize{\textbf{Training}}} 
\multirow{2}{*}{\textbf{Test Set}}
& \multicolumn{3}{c}{\textbf{Bovine Training Set,} \scriptsize{$\mathrm{Beef_{train}}$}}                  \\ 
\cmidrule(l){3-5} \cmidrule(l){7-9} 
\multicolumn{2}{c}{} & \multicolumn{1}{c}{\scriptsize{\textbf{Angle Error (°)}}} & \multicolumn{1}{c}{\scriptsize{\textbf{Tip Error (mm)}}} & \scriptsize{\textbf{TER (\%) $\downarrow$}} & 
& \multicolumn{1}{c}{\scriptsize{\textbf{Angle Error (°)}}} & \multicolumn{1}{c}{\scriptsize{\textbf{Tip Error (mm)}}} & \scriptsize{\textbf{TER (\%) $\downarrow$}} \\ 
\midrule
\multirow{3}{*}{UNet~\cite{ronneberger2015u}} & \scriptsize{$\mathrm{Pork_{test}}$} & 7.46$\pm$13.97 & 7.33$\pm$8.88 & 24.00 & \scriptsize{$\mathrm{Beef_{test}}$} & 6.62$\pm$11.58 & 7.37$\pm$7.04 & 27.81 \\
& \scriptsize{$\mathrm{Pork_{norm}}$} & 5.55$\pm$12.15 & 6.48$\pm$7.48 & 19.90 & \scriptsize{$\mathrm{Beef_{norm}}$} & 2.84$\pm$3.37 & 4.82$\pm$6.65 & 8.14 \\
& \scriptsize{$\mathrm{Pork_{chal}}$} & 9.29$\pm$15.30 & 8.15$\pm$9.98 & 27.94 & \scriptsize{$\mathrm{Beef_{chal}}$} & 10.49$\pm$15.17 & 9.97$\pm$6.44 & 47.75 \\ 
% \midrule
% \cmidrule(l){2-5} \cmidrule(l){6-9} 
\cmidrule(l){2-9} 
\multirow{2}{*}{WNet~\cite{chen2022automatic}} & \scriptsize{$\mathrm{Pork_{test}}$} & 5.86$\pm$14.01 & 5.21$\pm$6.41 & 24.16 & \scriptsize{$\mathrm{Beef_{test}}$} & 4.70$\pm$6.39 & 5.98$\pm$5.17 & 15.76 \\
& \scriptsize{$\mathrm{Pork_{norm}}$} & 3.09$\pm$7.23 & 3.74$\pm$4.47 & 14.68 & \scriptsize{$\mathrm{Beef_{norm}}$} & 4.77$\pm$8.12 & 6.05$\pm$6.34 & 16.39 \\
& \scriptsize{$\mathrm{Pork_{chal}}$} & 8.54$\pm$17.92 & 6.63$\pm$7.58 & 33.23 & \scriptsize{$\mathrm{Beef_{chal}}$} & 4.63$\pm$3.83 & 5.91$\pm$3.55 & 15.11 \\ 
% \midrule
\cmidrule(l){2-9} 
\multirow{2}{*}{VibNet (ours)} & \scriptsize{$\mathrm{Pork_{test}}$} & \textbf{1.45$\pm$1.49 }& \textbf{1.41$\pm$1.45} & \textbf{0.04} & \scriptsize{$\mathrm{Beef_{test}}$} & \textbf{1.72$\pm$4.98} & \textbf{1.35$\pm$1.82} & \textbf{0.84} \\
 & \scriptsize{$\mathrm{Pork_{norm}}$} & \textbf{1.25$\pm$0.91} & \textbf{1.20$\pm$1.30} & \textbf{0.00} & \scriptsize{$\mathrm{Beef_{norm}}$} & \textbf{1.26$\pm$0.96} & \textbf{1.22$\pm$1.33} & \textbf{0.00} \\
 & \scriptsize{$\mathrm{Pork_{chal}}$} & \textbf{1.64$\pm$1.86} & \textbf{1.61$\pm$1.56}  & \textbf{0.07} & \scriptsize{$\mathrm{Beef_{chal}}$} & \textbf{2.18$\pm$6.97} & \textbf{1.48$\pm$2.21}  & \textbf{1.69} \\
\bottomrule
\multicolumn{6}{l}{The angle error and tip error are described in the format of mean$\pm$std.}
\end{tabular}
}
\end{table*}

\subsection{Detection Performance Based on Needle Visibility}
\par
Accurately detecting the visibility-reduced or visually invisible needles, as shown in the challenging cases in Fig.~\ref{fig:usimage_visibility}, is a non-trivial task. Due to the existence of needle-like appearing anatomical artifacts, wrong predictions may result in severe damage to the surrounding vital tissue by further moving the needle forward. To better describe the needle detection performance, particularly for the invisible images, we introduce the evaluation metric called threshold exceedance rate (TER). TER measures the ratio of predictions with errors exceeding predefined thresholds: an angle error greater than $15^{\circ}$ or a tip error greater than $10$ mm in this study. It is important to note that these thresholds are not based on clinical acceptability but are used to demonstrate the robustness of the proposed method in challenging cases with reduced needle visibility. This is crucial for real-world applications where ground truth is unavailable.

Additionally, the traditional insertion angle error and needle tip position error (mean$\pm$std) are computed over all results for a fair comparison. This is because, in real-world scenarios, where the ground truth is unknown to the detection model, it is impossible for the model to determine if the results fall outside the TER threshold.
To compare the performance of the proposed VibNet with state-of-the-art (SOTA) methods, the most classical U-Net~\cite{ronneberger2015u} and recent W-Net~\cite{chen2022automatic} are employed in this study. The evaluation metrics include needle insertion angle error, needle tip position error, and TER. 

\begin{figure}[t]
    \centering
    \includegraphics[width=0.475\textwidth]{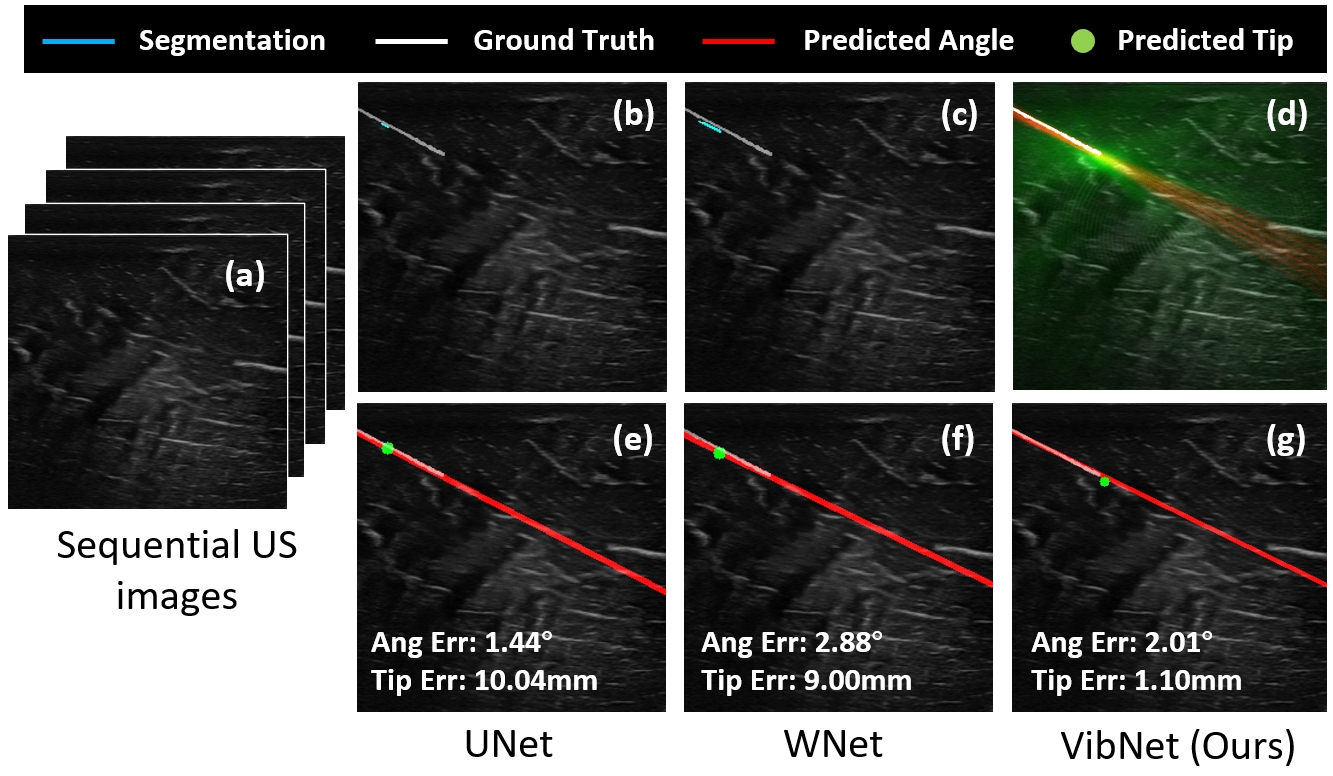}
    \caption{An example of detection results of UNet, WNet, and VibNet trained and tested on the bovine data. (a) shows the input US images where the needle is visually invisible. (b)-(d) are the intermediate results overlaid on the US image for intuitive visualization. In (d), the green and red colors indicate the probabilities of the predicted needle tip location and shaft direction, respectively. The final detection results are in (e)-(g). 
    }
    \label{fig:vis_res}
\end{figure}

\par
The needle detection results on the \textit{ex vivo} porcine and bovine tissues are depicted in the left and right parts, respectively, in Table~\ref{tab:comparison_seen}. For each tissue, the models of U-Net, W-Net, and VibNet were trained on the same training data set $\{\cdot\}_{train}$, and then the results were tested on unseen images from all test set $\{\cdot\}_{test}$, normal $\{\cdot\}_{norm}$ and challenging $\{\cdot\}_{chal}$ test sets (summarized in Table~\ref{tab:dataset}) to investigate the performance variance in terms of visibility. It can be seen from Table~\ref{tab:comparison_seen} that the proposed VibNet can significantly outperform the other two methods across all three metrics. For example, on Pork$_{test}$, the angle error, tip error, and TER are only $1.45\pm1.49^{\circ}$, $1.41\pm1.45~mm$ and $0.04\%$, while the ones obtained by UNet and WNet are $7.46\pm13.97^{\circ}$, $7.33\pm8.88~mm$ and $24\%$, and $5.86\pm14.01^{\circ}$, $5.21\pm6.41~mm$ and $24.16\%$, respectively. An intuitive visualization of the detection results for a randomly selected unseen image with reduced visibility from Beef$_{chal}$ using different methods is shown in Fig.~\ref{fig:vis_res}. VibNet achieves significantly higher tip accuracy ($1.10~mm$) compared to UNet ($10.04~mm$) and WNet ($9.0~mm$). Considering the desired needle tip detection performance in common procedures such as spine anesthesia ($< 5.0~mm$~\cite{greher2004ultrasound}) and genicular nerve radiofrequency ablation ($2.4\pm4.5~mm$~\cite{fonkoue2021validation}), the detection error $1.41 \pm 1.45~mm$ achieved by the proposed method is acceptable.

\par
In addition, we want to highlight the witnessed performance variation in terms of image visibility ($\{\cdot\}_{norm}$ vs. $\{\cdot\}_{chal}$). Both for porcine and bovine tissues, it can be seen from Table~\ref{tab:comparison_seen} that the significant performance drop on challenging cases in terms of all three metrics, e.g., angle error from $5.55\pm12.15^{\circ}$ to $9.29\pm15.30^{\circ}$, tip error from $6.48\pm7.48~mm$ to $8.15\pm9.98~mm$, and TER from $19\%$ to $27.94\%$, respectively, for porcine tissues using UNet. However, the proposed VibNet can result in comparable results for challenging data, e.g., $1.64\pm1.86^{\circ}$ vs. $1.25\pm0.91^{\circ}$, $1.61\pm1.56~mm$ vs. $1.20\pm1.30~mm$ and $0.07\%$ vs. $0$ in terms of three metrics on porcine data. This demonstrates the effectiveness of the proposed VibNet, which is less sensitive to the image intensity by leveraging the frequency information. This makes it possible for VibNet to detect the needle even when it is nearly invisible in US images. This enhancement improves the robustness of interventions and increases efficiency by reducing the need for reinsertion due to degraded needle visibility. More intuitive examples can be found in this video\footnote{The \textbf{video}: https://youtu.be/lXzHw0crPaM}.

\begin{table*}
\caption{Generalization Performance on Unseen Ex-vivo Tissues}
\label{tab:comparison_unseen}
\centering
\tabcolsep=0.20cm
\resizebox{0.90\textwidth}{!}{%
\begin{tabular}{lcccc|cccc}
\toprule
% Method & Test Dataset & Angle Error (°) & Tip Error (mm) & Outlier (\%) \\
\multirow{2}{*}{\textbf{Methods}} & 
% \multirow{2}{*}{\scriptsize{\textbf{Test Set}}{\Large$\backslash$}\scriptsize{\textbf{Training}}} 
\multirow{2}{*}{\textbf{Test Set}} 
& \multicolumn{3}{c}{\textbf{Porcine Training Set,} \scriptsize{$\mathrm{Pork_{train}}$}} & 
% \multirow{2}{*}{\scriptsize{\textbf{Test Set}}{\Large$\backslash$}\scriptsize{\textbf{Training}}} 
\multirow{2}{*}{\textbf{Test Set}}
& \multicolumn{3}{c}{\textbf{Bovine Training Set,} \scriptsize{$\mathrm{Beef_{train}}$}}                  \\ 
\cmidrule(l){3-5} \cmidrule(l){7-9} 
\multicolumn{2}{c}{} & \multicolumn{1}{c}{\scriptsize{\textbf{Angle Error (°)}}} & \multicolumn{1}{c}{\scriptsize{\textbf{Tip Error (mm)}}} & \scriptsize{\textbf{TER (\%) $\downarrow$}} & 
& \multicolumn{1}{c}{\scriptsize{\textbf{Angle Error (°)}}} & \multicolumn{1}{c}{\scriptsize{\textbf{Tip Error (mm)}}} & \scriptsize{\textbf{TER (\%) $\downarrow$}} \\ 
\midrule
\multirow{2}{*}{UNet~\cite{ronneberger2015u}} & \scriptsize{$\mathrm{Beef_{test}}$} & 11.03$\pm$18.43 & 12.75$\pm$10.40 & 52.40 & \scriptsize{$\mathrm{Pork_{test}}$} & 11.04$\pm$14.23 & 9.94$\pm$7.24 & 41.48 \\
& \scriptsize{$\mathrm{Beef_{norm}}$} & 4.17$\pm$4.30 & 9.39$\pm$8.48 & 35.58 & \scriptsize{$\mathrm{Pork_{norm}}$} & 8.70$\pm$14.67 & 7.47$\pm$6.70 & 26.02 \\
& \scriptsize{$\mathrm{Beef_{chal}}$} & 18.36$\pm$24.05 & 16.33$\pm$11.04 & 69.44 & \scriptsize{$\mathrm{Pork_{chal}}$} & 13.40$\pm$13.37 & 12.43$\pm$6.90 & 56.30 \\ 
% \midrule
% \cmidrule(l){2-5} \cmidrule(l){6-9} 
\cmidrule(l){2-9} 
\multirow{2}{*}{WNet~\cite{chen2022automatic}} & \scriptsize{$\mathrm{Beef_{test}}$} & 11.57$\pm$21.89 & 9.35$\pm$6.87 & 47.17 & \scriptsize{$\mathrm{Pork_{test}}$} & 13.62$\pm$21.63 & 9.18$\pm$8.68 & 38.77 \\
& \scriptsize{$\mathrm{Beef_{norm}}$} & 6.89$\pm$17.16 & 7.47$\pm$5.86 & 28.30 & \scriptsize{$\mathrm{Pork_{norm}}$} & 11.62$\pm$19.87 & 7.27$\pm$6.39 & 29.62 \\
& \scriptsize{$\mathrm{Beef_{chal}}$} & 17.43$\pm$25.48 & 11.71$\pm$7.31 & 66.30 & \scriptsize{$\mathrm{Pork_{chal}}$} & 15.76$\pm$23.19 & 11.22$\pm$10.21 & 47.53 \\ 
% \midrule
\cmidrule(l){2-9} 
\multirow{2}{*}{VibNet (ours)} & \scriptsize{$\mathrm{Beef_{test}}$} & \textbf{6.49$\pm$17.93} & \textbf{3.33$\pm$6.86} & \textbf{13.02} & \scriptsize{$\mathrm{Pork_{test}}$} & \textbf{1.83$\pm$2.56} & \textbf{1.22$\pm$1.07} & \textbf{0.39} \\
 & \scriptsize{$\mathrm{Beef_{norm}}$} & \textbf{1.35$\pm$1.47} & \textbf{1.46$\pm$3.40} & \textbf{1.03} & \scriptsize{$\mathrm{Pork_{norm}}$} & \textbf{1.33$\pm$1.28} & \textbf{1.03$\pm$0.66} & \textbf{0.00} \\
 & \scriptsize{$\mathrm{Beef_{chal}}$} & \textbf{11.69$\pm$24.33} & \textbf{5.23$\pm$8.71}  & \textbf{25.17} & \scriptsize{$\mathrm{Pork_{chal}}$} & \textbf{2.31$\pm$3.28} & \textbf{1.39$\pm$1.32}  & \textbf{0.77} \\
\bottomrule
\multicolumn{6}{l}{The angle error and tip error are described in the format of mean$\pm$std.}
\end{tabular}
}
\end{table*}

\subsection{Performance of Generalization Capability}
\par
\begin{figure}[!hbt]
    \centering
    \includegraphics[width=0.40\textwidth]{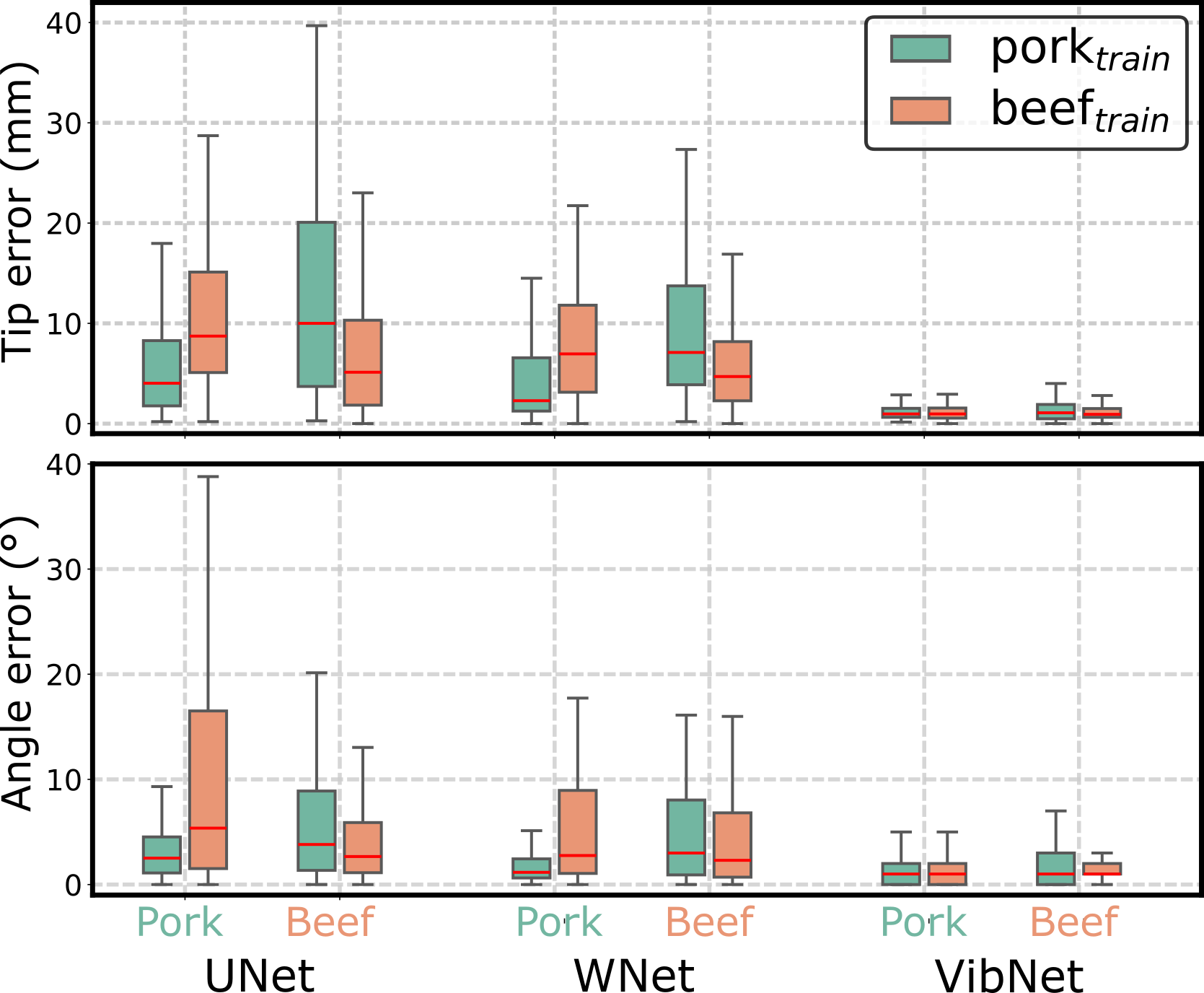}
    \caption{Statistical results obtained by applying trained models on two different ex-vivo tissues, each tested on both their own and the other type of animal tissues.}
    \label{fig:unseen_cmp}
\end{figure}

To further validate the adaptiveness of the method for different patients, we trained two models on the porcine and bovine training data separately and then tested these two models on the unseen animal samples. Since the VibNet relies more on the periodic motion pattern rather than the image intensity, the proposed VibNet is expected to maintain the performance when we test the model on the unseen type of animal samples. The angular and tip errors of the prediction results are summarized in Fig.~\ref{fig:unseen_cmp}. In all cases, certain performance drops are witnessed when we apply the trained models to an unseen type of animal sample. However, the extent of the decrease varies significantly among the three methods. Taking the porcine test data as an example, the tip and angle accuracy decreased significantly when applying UNet trained on bovine data compared to UNet trained on porcine data. Specifically, the tip error increased from $7.33\pm8.88~mm$ to $12.75\pm10.40~mm$, and the angle error increased from $7.46\pm13.97^{\circ}$ to $11.03\pm18.43^{\circ}$. But it is worth noting that such drops are much less for the proposed VibNet than others. It can be seen from Fig.~\ref{fig:unseen_cmp}, the VibNet model trained on Beef$_{train}$ achieves $1.35\pm1.82~mm$, and $1.72\pm4.98^{\circ}$ on Beef$_{test}$ and $1.22\pm1.07~mm$, and $1.83\pm2.56^{\circ}$ on Pork$_{test}$, respectively. The results on the unseen type of animal samples are comparable to the seen type of animal tissues. 

\par
To systematically validate the generalization performance of different methods on unseen ex-vivo samples, we further carried out experiments on normal ($\{\cdot\}_{norm}$) and challenging datasets ($\{\cdot\}_{chal}$). The results are summarized in Table~\ref{tab:comparison_unseen}. Similar to the conclusion obtained in the last section, significant performance drops are witnessed in all cases, from the normal dataset to the visibility-reduced dataset. Taking the model trained on Pork$_{train}$ as an example, it can be seen that UNet's performance on Beef$_{norm}$ decreased from $4.17\pm4.30^{\circ}$ to $18.36\pm24.05^{\circ}$ for angle error and from $9.39\pm8.48~mm$ to $16.33\pm11.04~mm$ for tip error on Beef$_{chal}$. However, it is worth noting that the proposed VibNet exhibits the most minor drops across all three metrics. Additionally, we noticed the significant performance variances between applying the model trained on Pork$_{train}$ to Beef$_{\{\cdot\}}$ (left half in Table~\ref{tab:comparison_unseen}) and using the model trained on Beef$_{train}$ on Pork$_{\{\cdot\}}$ (right part in Table~\ref{tab:comparison_unseen}). It can be seen from the table that the model trained on Beef data can achieve very close or even better results (tip error from $1.41\pm1.45~mm$ to $1.22\pm1.07~mm$) on porcine test data than that using the model trained on Pork data directly. This is because the learning-based periodic motion detection still relies on the image intensity variation along the temporal direction. In our case, the bovine tissue sample has fewer fat layers than the porcine tissue, thereby resulting in an overall lower contrast and brightness. The results demonstrate that VibNet is more robust than UNet and WNet because it leverages dynamic data, reducing the negative impact caused by acoustic shadows and linear-shaped echoes. 

\begin{table}
\caption{Detection Performance Based on Needle Insertion Angles}
\label{tab:comparison_angles}
\centering
\tabcolsep=0.20cm
\resizebox{0.475\textwidth}{!}{%
\begin{tabular}{lcccc}
\toprule
% Method & Test Dataset & Angle Error (°) & Tip Error (mm) & Outlier (\%) \\
\multirow{2}{*}{\textbf{Methods}} & 
% \multirow{2}{*}{\scriptsize{\textbf{Test Set}}{\Large$\backslash$}\scriptsize{\textbf{Training}}} 
\multirow{2}{*}{\textbf{Test Set}}
& \multicolumn{3}{c}{\textbf{Porcine Training Set,} \scriptsize{$\mathrm{Pork_{train}}$}} \\ 
\cmidrule(l){3-5}
\multicolumn{2}{c}{} & \multicolumn{1}{c}{\scriptsize{\textbf{Angle Error (°)}}} & \multicolumn{1}{c}{\scriptsize{\textbf{Tip Error (mm)}}} & \scriptsize{\textbf{TER (\%) $\downarrow$}} \\ 
\midrule
\multirow{3}{*}{UNet~\cite{ronneberger2015u}} & \scriptsize{$\mathrm{Pork_{shw}}$} & 1.50$\pm$1.28 & 4.23$\pm$7.26 & 11.42 \\
& \scriptsize{$\mathrm{Pork_{stp}}$} & 4.59$\pm$8.88 & 6.53$\pm$4.73 & 19.01 \\ 
% \midrule
% \cmidrule(l){2-5} \cmidrule(l){6-9} 
\cmidrule(l){2-5} 
\multirow{2}{*}{WNet~\cite{chen2022automatic}} & \scriptsize{$\mathrm{Pork_{shw}}$} & \textbf{1.01$\pm$0.64} & 2.15$\pm$2.13 & \textbf{0.00} \\
& \scriptsize{$\mathrm{Pork_{stp}}$} & 5.03$\pm$9.63 & 5.22$\pm$5.46 & 19.01 \\ 
% \midrule
\cmidrule(l){2-5} 
\multirow{2}{*}{VibNet (ours)} & \scriptsize{$\mathrm{Pork_{shw}}$} & 1.08$\pm$0.84 & \textbf{0.81$\pm$0.34} & \textbf{0.00} \\
& \scriptsize{$\mathrm{Pork_{stp}}$} & \textbf{1.47$\pm$0.95} & \textbf{1.06$\pm$0.46}  & \textbf{0.00} \\
\bottomrule
% \multicolumn{6}{l}{\scriptsize PTa/PTA and BTa/BTA are defined in Table \ref{tab:dataset}.}
\multicolumn{5}{l}{The angle error and tip error are described in the format of mean$\pm$std.}
\end{tabular}
}
\end{table}

\subsection{Performance Variation Based on Insertion Angle}
\par
Since the needle insertion angle is commonly considered a key factor that will significantly affect the detection accuracy, this section validates the performance of the proposed VibNet with data acquired at both low ($\le 30^{\circ}$) and steep insertion angles ($>30^{\circ}$). Based on previous results, the performance of the models across porcine and bovine data was consistent. For simplicity, we use only porcine data to illustrate how insertion angle affects performance. The comparison of UNet, WNet, and VibNet is summarized in Table~\ref{tab:comparison_angles}. It can be seen that the performance in terms of all three metrics is consistently better on the data with a low insertion angle than the ones with a steep angle across all three models. This is because low-angle insertion usually has better needle visibility than steep insertion due to the high amount of echos reflected back to the transducer. For Pork$_{shw}$, the angle errors obtained by the three methods are close to each other (VibNet: $1.08\pm0.84^{\circ}$ vs. UNet: $1.5\pm1.28^{\circ}$ and WNet: $1.01\pm0.64^{\circ}$), 
% ($1.5\pm1.28^{\circ}$ vs. $1.01\pm0.64^{\circ}$ vs. $1.08\pm0.84^{\circ}$), 
while the tip error obtained by the proposed VibNet ($0.81\pm0.34~mm$) performs much better than UNet ($4.23\pm7.26~mm$) and WNet ($2.15\pm2.13~mm$).

\begin{table*}
\caption{Ablation Study Results}
\label{tab:ablation}
\centering
\tabcolsep=0.20cm
\resizebox{0.98\textwidth}{!}{%
\begin{tabular}{lccccc|ccccc}
\toprule
% Method & Test Dataset & Angle Error (°) & Tip Error (mm) & Outlier (\%) \\
\multirow{2}{*}{\textbf{Frameworks}} & \multicolumn{5}{c}{\textbf{Porcine Training/Test Sets:} \scriptsize{$\mathrm{Pork_{train}}$~/~$\mathrm{Pork_{test}}$}} & \multicolumn{5}{c}{\textbf{Bovine Training/Test Sets:} \scriptsize{$\mathrm{Beef_{train}}$~/~$\mathrm{Beef_{test}}$}}\\ 
\cmidrule(l){2-6} \cmidrule(l){7-11} 
\multicolumn{1}{c}{} & \multicolumn{1}{c}{\scriptsize{\textbf{Angle Error (°)}}} & \scriptsize{\textbf{$p_{ang}$-Val.}} & \multicolumn{1}{c}{\scriptsize{\textbf{Tip Error (mm)}}} & \scriptsize{\textbf{$p_{tip}$-Val.}} & \scriptsize{\textbf{TER (\%) $\downarrow$}} & \multicolumn{1}{c}{\scriptsize{\textbf{Angle Error (°)}}} & \scriptsize{\textbf{$p_{ang}$-Val.}} & \multicolumn{1}{c}{\scriptsize{\textbf{Tip Error (mm)}}} & \scriptsize{\textbf{$p_{tip}$-Val.}} & \scriptsize{\textbf{TER (\%) $\downarrow$}} \\ 
\midrule
VibNet$^{*-}$ & 8.03$\pm$11.46 & $<0.001$ & $N/A$ & $N/A$ & 18.65 & 13.17$\pm$17.85 & $<0.001$ & $N/A$ & $N/A$ & 31.62 \\ 
VibNet w/o Enc. init. & 3.28$\pm$9.50 & $<0.001$ & 1.75$\pm$2.32 & $<0.001$  & 4.25  & 2.07$\pm$5.63 & $<0.001$  & 1.44$\pm$2.46 & 0.013  & 2.13 \\
VibNet w/o STFT init. & 1.65$\pm$2.54 & $<0.001$  & 1.47$\pm$1.59 & 0.013  & 0.38  & 3.52$\pm$8.15 & $<0.001$  & 2.30$\pm$4.02 & $<0.001$  & 11.06 \\
VibNet rp. DHT & 1.74$\pm$2.85 & $<0.001$  & 1.68$\pm$1.80 & $<0.001$  & 2.13  & 2.01$\pm$2.51 & $<0.001$  & 1.54$\pm$3.16 & $<0.001$  & 1.72 \\ 
VibNet w. BCE loss & \textbf{1.17$\pm$1.09} & $<0.001$  & \textbf{1.35$\pm$1.64} & 0.012  & \textbf{0.0} & 2.20$\pm$3.68 & $<0.001$  & 1.92$\pm$3.71 & $<0.001$  & 5.87 \\ 
% VibNet$^{*-}$ &  & 8.03$\pm$11.46 & $N/A$ & 18.65 &  & 13.17$\pm$17.85 & $N/A$ & 31.62 \\ 
\midrule
${^{\dagger}}$UNet30 & 13.31$\pm$23.70 & $<0.001$  & 10.43$\pm$10.57 & $<0.001$  & 44.11 & 9.64$\pm$19.70 & $<0.001$  & 8.09$\pm$8.15 & $<0.001$  & 39.14 \\ 
${^{\dagger}}$UNet3D + LSTM~\cite{belikova2021deep} & 7.94$\pm$16.70 & $<0.001$  & 7.55$\pm$8.46 & $<0.001$  & 24.25 & 17.80$\pm$24.93 & $<0.001$  & 12.06$\pm$9.62 & $<0.001$  & 48.79 \\ 
\midrule
VibNet (ours) & 1.45$\pm$1.49 & -  & 1.41$\pm$1.45 & -  & 0.04  & \textbf{1.72$\pm$4.98} & -  & \textbf{1.35$\pm$1.82} & -  & \textbf{0.84} \\ 
\bottomrule
\multicolumn{11}{l}{\scriptsize{The angle error and tip error are described in the format of mean$\pm$std,~~ w.: with, ~~w/o: without,~~~rp.: replace, ~~~${*-}$: rp. all learnable modules, ~~~$N/A$: not applicable, ~~$\dagger$: 30 US images as input}.} \\
\multicolumn{11}{l}{\scriptsize{$p_{ang}$-Val. and $p_{tip}$-Val. denote the $p$-values for the angle and tip errors, respectively, computed from the prediction errors of VibNet and the compared method. 
% ***: $p<0.001$, *: $0.01\le p\le 0.05$
}}
\end{tabular}
}
\end{table*}

\par
For steep insertion angles, the proposed VibNet outperforms all other methods in three key metrics: angle error ($1.47\pm0.95^\circ$), tip error ($1.06\pm0.46~mm$), and TER (0). It also experiences the smallest performance drop compared to other methods, with reductions of $0.39^\circ$ in angle error and $0.25~mm$ in tip error. Thus, the frequency feature-based VibNet demonstrates superior robustness and potential for applications requiring steep insertion angles compared to the intensity-based UNet and WNet.

\subsection{Ablation Study}
\par
To validate the impact of each key component of VibNet, we conducted ablation studies to quantify the contributions of each module. We retrained VibNet under different configurations to analyze their effects:
% \begin{enumerate}
\begin{itemize}%[label=--]
    \item \textit{w/o Enc. init.}: The Encoder was randomly initialized to validate its effectiveness in characterizing the subtle motion of the needle,
    \item \textit{w/o STFT init.}: The STFT module was randomly initialized to demonstrate the improvement gained by converting the extracted features from the temporal domain to the frequency domain,
    \item \textit{rp. DHT}: The DHT module was replaced with a CNN-based segmentation header to highlight the necessity of predicting the needle in Hough space,
    \item \textit{w. BCE loss}: The Focal loss was replaced with a BCE loss for comparison.
\end{itemize}
% \end{enumerate}
Furthermore, to emphasize the importance of integrating learnable modules, we constructed a needle detection framework denoted as ``VibNet$^{*-}$''. In this variant, all learnable modules in VibNet were sequentially replaced with Riesz pyramid \cite{wadhwa2014riesz}, bandpass filters, and Hough transform. The results on both porcine and bovine data are summarised in Table \ref{tab:ablation}. Notably, ``VibNet$^{*-}$'', which excludes any learnable modules, performs the worst with the highest TERs and angle errors, e.g., angle error 13.17$\pm$17.85$^\circ$ on Beef$\mathrm{_{test}}$. This is because the traditional feature-based method can not distinguish the needle itself and the surrounding tissues with similar motion patterns. Therefore, even with accurate motion area extraction, detecting the needle, particularly the needle tip, is inherently challenging. Furthermore, the customized feature filters usually lack adaptiveness to various images on unseen images. To demonstrate the statistical significance of the results achieved by the proposed VibNet and other architectures, we also compute the $p$-value between each pair using the two-tailed paired \textit{t}-test \cite{orlando2023power}. The the results ($p_{ang}$ and $p_{tip}$) are summarized in Table~\ref{tab:ablation}.

\par
Table~\ref{tab:ablation} shows that the detection performance across all three metrics for both porcine and bovine tissues degrades when the encoder and/or STFT module in VibNet are not initialized. This highlights the importance of proper initialization for achieving optimal results. The biggest drops occur on Beef$_{test}$ in terms of TER from $0.84\%$ to $11.06\%$ when Fouier initialization is removed (w/o STFT init.). Additionally, notable performance variations were observed across different types of ex-vivo tissues. This indicates that reasonable initialization is beneficial for encouraging the network's convergence, which is crucial when optimizing multiple modules in an end-to-end manner. In terms of the loss function, using BCE loss marginally outperforms the proposed framework on Pork$_{test}$ but results in a noticeable performance decrease on Beef$_{test}$ (TER 5.87\%). The results demonstrate that using Focal Loss results in more consistent outcomes with lower TER, particularly for bovine tissue where overall brightness is low. Therefore, we adopted Focal Loss to train the final VibNet. We also investigated the performance of replacing the DHT module with an additional CNN-based segmentation layers (rp. DHT). This modification led to a decrease in overall performance across all three metrics: angle error increased from $1.45\pm1.49^{\circ}$ to $1.74\pm2.85^{\circ}$ ($p_{ang}<0.001$), tip error increased from $1.41\pm1.45$~mm to $1.68\pm1.80$~mm ($p_{tip}<0.001$), and TER increased from $0.04\%$ to $2.13\%$ on Pork$_{test}$, as shown in Table~\ref{tab:ablation}.
Notably, all $p$-values comparing the proposed method with others are below $0.05$ for both tissue types in terms of angular and tip error (see Table~\ref{tab:ablation}). This indicates that VibNet achieves statistically significant improvements over its peers.

To provide a more comprehensive evaluation of VibNet’s ability to handle temporal information, we compare VibNet with a modified UNet structure~\cite{ronneberger2015u} (referred to as UNet30), which takes 30 concatenated US images as input, and a UNet3D+LSTM model~\cite{belikova2021deep}, which incorporates LSTM layers to process temporal features and also uses 30 US images as input for prediction. The experimental results presented in Table~\ref{tab:ablation} show that neither increasing the length of the input sequence nor incorporating an LSTM layer into the classical architecture improves performance compared to the standard UNet. For example, tip errors on the Pork$_{test}$ are $10.43\pm10.57$~mm for UNet30, $7.55\pm8.46$~mm for UNet3D+LSTM, and $7.33\pm8.88$~mm for the standard UNet. This may be because, with longer input sequences, the models relying on the pixel intensities for prediction are more prone to overfitting texture features, even when temporal features are considered (see the results for $^\dagger$UNet+LSTM in Table~\ref{tab:ablation}). Such overfitting likely causes the models to struggle with inputs outside the distribution of the training data, leading to degraded performance. 

\par
To demonstrate the time efficiency of the proposed VibNet, the average time consumption for each module (as shown in Fig. \ref{fig:network_overview}) was calculated using 1000 input samples. The results are summarized in Table \ref{tab:time_collapse}. Notably, the CNN-based temporal feature extraction and frequency feature aggregation significantly enhance time efficiency compared to traditional filters~\cite{wadhwa2013phase}, requiring only $0.7 \pm 0.1~ms$ and $0.6 \pm 0.0~ms$, respectively. However, the deep Hough transformation module and post-processing for needle shaft and tip determination take $49.2 \pm 0.4~ms$ and $33.7 \pm 0.3~ms$, respectively. Overall, the total time consumption is approximately $84.2~ms$, resulting in an inference frequency of around 12 Hz. This relatively high inference time will limit its application for high-dynamic interventions, such as cardiac application, while it is sufficient for low-dynamic needle interventions with steady, slow movements, such as nerve blocks and biopsies.

\begin{table}[!h]
\caption{Inference Time of VibNet in Fig. \ref{fig:network_overview} (mean$\pm$std)}
\label{tab:time_collapse}
\centering
\tabcolsep=0.20cm
\resizebox{0.485\textwidth}{!}{%
\begin{tabular}{ccccc}
\toprule
\textbf{Module (a)} & \textbf{Module (b)} & \textbf{Module (c)} & \textbf{Module (d)} & \textbf{Total} \\
\midrule
0.7$\pm$0.1~ms & 0.6$\pm$0.0~ms & 49.2$\pm$0.4~ms & 33.7$\pm$0.3~ms & 84.2$\pm$0.5~ms \\
% \midrule
\bottomrule
% \multicolumn{4}{l}{\scriptsize \textbf{Unit}: MS, MEAN$\pm$STD.}
\end{tabular}
}
\end{table}

\section{Discussion}
\label{sec:discussion}
\par
This study first presents a deep-learning-based network, VibNet, to address the challenge of detecting inserted needles when their visibility is reduced or nearly invisible in US images. Based on the experiments carried out on two different types of ex-vivo samples, VibNet can provide robust and accurate detection results of the inserted needles by exploring the features in the frequency domain instead of traditional image intensity. To the best of our knowledge, this is the first end-to-end network leveraging vibration to do needle segmentation. It can be integrated into the current clinical routine by introducing a motorized mechanism generating periodic subtle vibrations externally to help clinicians precisely locate the percutaneous needle tip when it is not recognizable by the naked eye in images. This integration enhances safety, and increases time efficiency by reducing the need for re-insertion due to the loss of needles in images. The current approach focuses on in-plane applications for 2D images, which are commonly used in current practice. For out-of-plane applications, where the needle appears as a single point, performance decreases due to: 1) relatively large manual annotation errors for single intersection points, and 2) the absence of joint geometric constraints between needle shaft and tip in Hough space, leading to more outliers in detection results. However, we emphasize that the method remains effective even with minor misalignments due to a specific artifact known as ``slice thickness"~\cite{jiang2024needle, goldstein1981slice}. This artifact results in a line-like feature rather than a single intersection point in the final 2D images. 

\begin{figure}[t]
    \centering
    \includegraphics[width=0.485\textwidth]{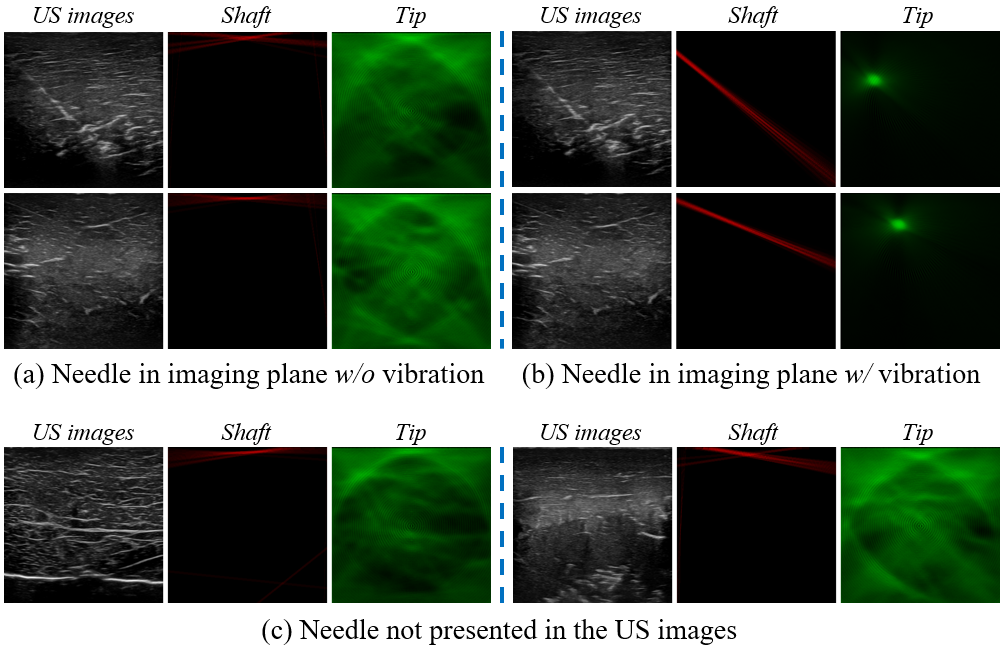}
    \caption{VibNet's detection performance under specific conditions.}
    \label{fig:vis_examples}
\end{figure}

\par
A linear probe was used in this study to validate the concept of vibration-based needle segmentation, as it allows ground truth generation for needles with reduced visibility by manually adjusting the steering angle. In clinical practice, linear probes are commonly used for visualizing shallow tissues, such as cervical facet joint injections~\cite{bodor2022ultrasound}, while convex probes are preferred for deep tissues. Both are widely utilized. Since VibNet mainly processes images temporally across multiple frames in terms of pixel location, its performance is not dependent on the US image shapes. So, the proposed method is also applicable to convex probes.

The current work primarily focuses on the feasibility of leveraging vibration using the network and investigating the potential of the novel frequency-based method over traditional intensity-based methods. We validated the performance across different types of animal tissues for generalization. However, further investigation is needed to address practical factors in real patients, such as insertion depth, needle bending, needle types, patient motion~\cite{jiang2022towards}, and probe-induced tissue deformation~\cite{jiang2021deformation}, to bring VibNet into clinical practice.

\par
The VibNet architecture leverages frequency information from consecutive images to address challenges with reduced instrument visibility, but its performance declines without periodic vibration. This is evident in Fig.~\ref{fig:vis_examples} (a), where VibNet fails to detect the needle when vibration is off. In the same scene with vibration, VibNet accurately identifies both the shaft and tip [see Fig.~\ref{fig:vis_examples} (b)]. This outcome aligns with our design objective: VibNet learns motion patterns by analyzing pixel intensity variations across frames rather than relying solely on image intensity. Similarly, when the needle is absent in the US images, VibNet fails to detect it [see Fig.~\ref{fig:vis_examples} (c)] due to the absence of line-like features at the expected frequency.

\par
In this study, vibration is introduced when needle visibility is reduced in US images, targeting the practical challenge of visibility reduction during interventions. While VibNet performed well on both good and poor visibility cases (see Table~\ref{tab:comparison_angles}), the data used was recorded without insertion motion. For continuous tool monitoring, further investigation including insertion motion is needed. Besides, thinner or more flexible needles could reduce tissue displacement during vibration, which may potentially affect the final performance. Future work will focus on addressing these practical challenges in real scenarios, systematically analyzing the relationship between model performance and insertion angles, and will investigate the most effective vibration pattern, including different frequencies and amplitudes, for different clinical applications. 

\section{Conclusion}
\label{sec:conclusion}
\par
In this paper, we presented VibNet, a deep learning-based framework designed to enhance needle detection in US images by leveraging mechanism-induced vibrations. The model was evaluated using datasets collected from two distinct tissue samples, demonstrating its effectiveness in handling variations across different tissues. Compared to conventional intensity-based solutions such as UNet and the recent WNet, VibNet excels at extracting and utilizing robust frequency features from image sequences, leading to superior prediction results, particularly for needle tip detection. Even when needle visibility is reduced or the needle is nearly invisible, VibNet can robustly and accurately detect the needle tip ($1.61\pm1.56~mm$ compared to $8.15\pm9.98~mm$ for UNet and $6.63\pm7.58~mm$ for WNet on Pork$_{chal}$) and the needle direction ($1.64\pm1.86^{\circ}$ compared to $9.29\pm15.30^{\circ}$ for UNet and $8.54\pm17.92^{\circ}$ for WNet). We believe this study will pave the way for advanced needle detection in US procedures by offering seamless integration and enhanced robustness across various tissues. In addition, the core idea of this study is to detect objects in the frequency domain, offering a promising alternative for addressing dynamic and practical challenges in developing future autonomous intervention systems. This approach can also be extended to other scenarios with limited or reduced visibility, such as tool detection in laparoscopy images.

%%%%%%%%%%%%%%%%%% main content

\section*{ACKNOWLEDGMENT}
This study was supported in part by the Multi-scale Medical Robotics Center, AIR@InnoHK, Hong Kong, and the SINO-German Mobility Project (M0221). This article reflects the authors’ opinions and conclusions and not any other entity. In addition, the authors would like to acknowledge the editors and volunteering reviewers for their time and implicit contributions to the improvement of the article’s thoroughness, readability, and clarity.

\bibliographystyle{IEEEtran}
\bibliography{IEEEabrv, reference}

\end{document}